\documentclass[aps,prb,reprint,superscriptaddress,amsmath]{revtex4-1}
\usepackage{graphicx}
\usepackage{enumerate}
\usepackage{refcount}
\usepackage{fmtcount}
\usepackage{hyperref}

\hypersetup{pdfstartview=FitH,pdfpagemode=UseOutlines,colorlinks,
citecolor=blue,linkcolor=blue,urlcolor=blue,breaklinks}

\renewcommand{\a}{\alpha}
\renewcommand{\b}{\beta}
\newcommand{\g}{\gamma}
\renewcommand{\d}{\delta}

\newcommand{\s}{\sigma}
\renewcommand{\t}{\theta}
\newcommand{\dg}{\dagger}
\newcommand{\parDiag}{{\!/\!/\!}}

\begin{document}

%\preprint{}

\title{Electrical and thermal transport in the quasi-atomic limit of coupled Luttinger liquids}

\author{Aaron Szasz}
	\email[]{aszasz@berkeley.edu}
	\affiliation{Department of Physics, University of California, Berkeley, California 94720, USA}
\author{Roni Ilan}
	\affiliation{Department of Physics, University of California, Berkeley, California 94720, USA}
	\affiliation{Raymond and Beverly Sackler School of Physics and Astronomy, Tel Aviv University, Tel Aviv 69978, Israel}
\author{Joel E. Moore}
	\affiliation{Department of Physics, University of California, Berkeley, California 94720, USA}
	\affiliation{Materials Sciences Division, Lawrence Berkeley National Laboratory, Berkeley, California 94720, USA}

\date{\today}

\begin{abstract}
We introduce a new model for quasi one-dimensional materials, motivated by intriguing but not yet well-understood experiments that have shown two-dimensional polymer films to be promising materials for thermoelectric devices.  We consider a two-dimensional material consisting of many one-dimensional systems, each treated as a Luttinger liquid, with weak (incoherent) coupling between them.  This approximation of strong interactions within each one-dimensional chain and weak coupling between them is the ``quasi-atomic limit.''  We find integral expressions for the (interchain) transport coefficients, including the electrical and thermal conductivities and the thermopower, and we extract their power law dependencies on temperature. Luttinger liquid physics is manifested in a violation of the Wiedemann-Franz law; the Lorenz number is larger than the Fermi liquid value by a factor between $\gamma^2$ and $\gamma^4$, where $\gamma\geq 1$ is a measure of the electron-electron interaction strength in the system.
\end{abstract}

%\pacs{71.10.Fd, 71.10.Pm, 72.80.Le, 73.50.Lw, 73.61.Ph}

\maketitle

%=======================================================================
%=======================================================================

\section{Introduction\label{section:introduction}}

Recent experiments on thin films of doped polymers such as PEDOT-PSS, PEDOT-Tos, and PBTTT have found both high conductivity and a large thermopower \cite{Yuen2009,Pipe2013,Bubnova2013,Chabinyc2015}.  There are some possible explanations for the source of conductive behavior in polymers \cite{Heeger1988,Sai2007}, but they are not yet definitive; in this paper we will bypass this question and instead analyze a different facet of the problem.  Namely, since we take for granted the conductive nature of individual polymers, we can describe each polymer as a Luttinger liquid, and we look for possible signatures of the Luttinger liquid behavior that survive even when the one-dimensional systems are coupled to form a quasi-two-dimensional material.

The Luttinger liquid model \cite{Haldane1981Luttinger,GiamarchiBook} represents a one-dimensional electron gas modified by interaction between the electrons and can therefore be viewed as the one-dimensional analogue of the more well-known Fermi liquid model, though the generic behavior of the system is quite different.  In the Fermi liquid theory for two- and three-dimensional systems, the interacting system actually behaves very much like the corresponding non-interacting electron gas---the excitations are fermionic quasiparticles which behave qualitatively like electrons even if specific properties like mass are renormalized to new values.

By contrast, in one dimension interactions between electrons have a strong qualitative effect on the behavior of the system.  Schematically, one can picture electrons in higher-dimensional systems having space to ``go around'' each other and thus they still remain roughly independent (noninteracting), while in one dimension this is impossible, and so electrons will move together, forming collective (bosonic) excitations.  The Luttinger liquid theory and the technique of bosonization make this intuitive idea concrete.  

There are numerous convincing experimental results on one-dimensional systems that confirm various predictions of the Luttinger liquid theory.  For instance, Luttinger liquids are expected to exhibit spin-charge separation, where charge and spin degrees of freedom act independently \cite{GiamarchiBook}; spin-charge separation has been convincingly observed via photoemission experiments in artificially created one-dimensional structures \cite{Segovia1999}.  Likewise, the density of states around the Fermi level is predicted to show a distinctive power law behavior \cite{Braunecker}; this was also observed in an artificially created 1D chain \cite{Blumenstein2011}.  Other observations of Luttinger liquid-like behavior, however, have been made not on actual one-dimensional chains but rather on two-dimensional collections of one-dimensional systems such as in the polymer films that motivated this work\cite{Yuen2009} or on highly anisotropic three-dimensional crystals\cite{Moser1998,Wakeham2011,Bangura2013}; it is not immediately clear that the results of these experiments should be directly compared to theories of single Luttinger liquids.  Rather, the coupling of 1D chains to form a quasi-2D material may modify or destroy altogether the distinctive signatures of Luttinger liquid behavior.  A theory of coupled Luttinger liquids would thus be very helpful. 

While the theory of weakly coupled Luttinger liquids has been considered in the past by many different authors, there are very few results for thermal transport in a system of infinitely many coupled chains.  Some results deal with ``ladders'' consisting of just two coupled chains \cite{ClarkeStrong1996,GiamarchiBook}, while some of the most well-known treat coupling two half-infinite chains at their ends as a way of modeling an impurity in the Luttinger liquid\cite{KaneFisher1992,KaneFisher1992b,KaneFisher1996}.  Both electrical and thermal transport have also been computed for many impurities on a single chain\cite{Li2002}.  Papers that do consider an infinite array of weakly coupled Luttinger liquids have mostly focused only on the electrical conductivity \cite{Moser1998,Anderson1995,Giamarchi2000_2013} and not on any kind of thermal transport.  There is one recent paper on the off-diagonal terms of the thermopower tensor for infinitely many coupled chains\cite{Schattner2016Arxiv}, but we are not aware of any existing results for the thermal conductivity or Lorenz number in the type of model we consider.  This is the gap we intend to fill.

In this work we consider a model of coupled one-dimensional systems in which each 1D chain is treated as a (spinless) Luttinger liquid, and the individual chains are coupled by a perturbatively weak interchain hopping.  We refer to this situation of strong interactions within 1D chains and weak, incoherent coupling between them as the ``quasi-atomic limit.''  The approximations and assumptions inherent in this model, as well as some justifications of their validity, are discussed in section \ref{section:assumptions}.

We consider two somewhat different versions of the model, which incorporate Luttinger liquid behavior at different stages of the calculation.  In both cases, we calculate transport coefficients using the Kubo formalism.  In the first model, discussed in section \ref{section:nonint_model}, the electronic system is initially assumed to be noninteracting so that the state of the system can be described by occupation of single-particle orbitals; we introduce Luttinger behavior via the electronic density of states.  In the second model (section \ref{section:LL}), we use the full Luttinger liquid correlation functions.  In section \ref{section:discussion}, we summarize our key results and their applicability to experimental systems, and we further discuss the comparison between the two models.

We find that both models predict the same power law dependence on temperature for the transport coefficients, $\sigma \propto T^{2\g-3}$ and $\kappa \propto T^{2\g-2}$, where $\g$ is a measure of interaction strength as defined in equation \eqref{eq:def_gamma}, but that the precise values of the transport coefficients (as measured by the Lorenz number) vary with electron-electron interaction strength more strongly in the second, more complete, calculation.  In the generalized noninteracting model (section \ref{section:nonint_model}) we find that the Lorenz number is larger than the value predicted by the Wiedemann-Franz law by a factor between $\g^2$ and $\g^{2.4}$.  In the full Luttinger liquid model (section \ref{section:LL}), we find an even larger violation, with the Lorenz number augmented by as much as $\g^{3.6}$.

%=======================================================================
%=======================================================================

\section{Assumptions and approximations: the quasi-atomic limit\label{section:assumptions}}

In the Hubbard model, the ``atomic limit'' is the limit as the hopping between lattice sites vanishes while electron-electron interaction is held constant \cite{Beni1974,Mukerjee2005}.  We study the problem of weakly coupled chains with a similar approach, in which we do a perturbative calculation to lowest order in the interchain hopping while treating each one-dimensional chain as a single coherent quantum system.  This limit of full coherence in one direction (along chains) and weak incoherent hopping in the other direction (between chains) we call the ``quasi-atomic limit.''\footnote{Note that the term ``quasi-atomic limit'' has been used in the past to describe situations between full coherence and the atomic limit\cite{Siringo1991,Rojas2012}; we use it instead to indicate a system that is fully in the atomic limit in one direction and not at all in the other.}

To be more precise, we make the following assumptions:
\begin{enumerate}[(1)]
\item \label{assumption:independent_polymers}There is no electron-electron interaction between the 1D chains.
\item \label{assumption:atomic_limit}The different chains are perturbatively coupled through a weak hopping of electrons between adjacent chains.
\item \label{assumption:polymer_array}The 1D chains are located at evenly spaced points along a one-dimensional line, meaning that electrons may hop from one polymer to adjacent ones on either side of it and that the hopping strength between any pair of adjacent polymers is the same.
\label{assumption:final}
\end{enumerate}

\noindent We will briefly justify the applicability of these assumptions to real physical systems, beginning with assumption \eqref{assumption:atomic_limit}.  To measure transport properties for a macroscopic object (like a polymer film) we really want to use not the microscopic model of the system but rather an effective theory that results from a renormalization group flow.  At zero temperature, any coupling between chains is a relevant perturbation in the renormalization group sense, but fortunately this is not the case at finite temperature\cite{Anderson1994,ClarkeStrong1996,Biermann2002}.  This means that, so long as the temperature is much higher than the energy scale of the interchain coupling, the atomic limit will be valid.  For any particular material, this sets a lower bound on the temperature regime in which our results are applicable.

In this temperature regime of validity, the thermalization time within each chain (proportional to $1/T$) will be much less than the interchain hopping time (proportional to the inverse hopping strength), so that each individual one-dimensional chain will thermalize between hopping events.  We can therefore intuitively think of the interchain hopping as incoherent, though we do not explicitly use that fact anywhere in our calculations.

Assumption \eqref{assumption:polymer_array} is an accurate description for the case of anisotropic crystals.  The application to polymer films is less direct, as they are known to have regions where the polymers are relatively aligned in some organized array (as in assumption \ref{assumption:polymer_array}), as well as amorphous regions \cite{Stanford_experiment_paper,Bubnova2013,Steyrleuthner2014}. In the latter regions, which may account for a significant fraction of the overall film, as long as the polymers form a single two-dimensional layer and do not cross, at a sufficiently small scale the polymers should still form a neat array and our assumption will apply.  We can therefore approximately treat the film as consisting of a collection of randomly oriented domains, each of which individually satisfies the assumption.  We discuss this further in section \ref{section:discussion}.  

%=======================================================================
%=======================================================================

\section{Generalized noninteracting model\label{section:nonint_model}}

The first version of our model is intended to capture the key Luttinger liquid behavior while still being simple enough to provide helpful physical intuition about the system we study.  We thus use a noninteracting  model for most of the calculation, finally substituting the Luttinger liquid density of states at the end.  

To be precise, we add two more simplifying assumptions to those given in section \ref{section:assumptions} above:
\begin{enumerate}[(1)]
\setcounter{enumi}{3}%{\ref{assumption:final}}
\item \label{assumption:noninteracting}Each individual 1D chain can be described by a set of non-interacting single-particle orbitals, given by the Fourier modes of the localized on-site orbitals; the orbitals' energies are distributed according to the tunneling density of states of a Luttinger liquid, and each chain's orbitals are the same.
\item \label{assumption:hopping}Electrons hop from a well-defined single-particle eigenstate on one chain to an eigenstate with approximately the same energy and momentum on an adjacent chain.  The hopping strength is sharply peaked in $|k-k'|$ where $k$ and $k'$ are the wavenumbers on the two chains, and the value at $k=k'$ is independent of $k$.  (In practice, we assume the hopping is Gaussian in $k-k'$, but this assumption is only needed when we compare the two versions of our model, see Appendix \ref{appendix:correspondence}.)
\label{assumption:nonint_final}
\end{enumerate}

\noindent The \numberstringnum{\getrefnumber{assumption:nonint_final}} assumptions above lead to a specific interpretation of the standard tight-binding Hamiltonian
\begin{equation}
H = \sum_{j,k} E_k c^\dg_{jk} c_{jk} - \sum_{jkk'} \left(t_{kk'}c^\dg_{j,k} c_{j+1,k'} + \text{h.c.}\right)\label{eq:nonint_H}
\end{equation}

\noindent The index $j$ labels 1D chains, while $k$ and $k'$ label extended (Fourier state) orbitals on each chain.  $c^\dg$ and $c$ are the usual fermion creation and annihilation operators, while $E_k$ is the single-particle energy corresponding to the orbital $k$.  

In the noninteracting limit, the $E_k$ are just the energies of a one-dimensional tight-binding model $H_0 = -t_\parDiag\sum_i c^\dg_i c_{i+1} +\text{h.c.}$; if the lattice spacing is $a$, the energy levels are $E_k = -2t_\parDiag\cos(ka)$, which are then linearized around the Fermi points $k=\pm k_F$.  When interactions are introduced, there are no longer well-defined single-particle orbitals, so we cannot give an explicit formula for the energies $E_k$.  Instead, we will derive an expressions for the transport coefficients in which the energy spectrum only appears via the density of states, for which we can use the well-defined single-particle tunneling density of states of a Luttinger liquid.

\subsection{Calculation of transport coefficients}

We calculate the transport coefficients in this model using the Kubo formalism.  For consistency with standard references, we use the conventions of reference \onlinecite{Mahan3},  in which case the electrical conductivity, thermal conductivity, and thermopower are given by
\begin{subequations}
\begin{align}
\sigma & = \frac{e^2}{T}L^{(11)}\label{eq:sigma_formula}\\
\kappa & = \frac{1}{T^2}\left[L^{(22)} - \frac{(L^{(12)})^2}{L^{(11)}}\right]\label{eq:kappa_formula}\\
S & = - \frac{1}{eT}\frac{L^{(12)}}{L^{(11)}}\label{eq:S_formula}
\end{align}
\label{eq:coeffs_from_L}
\end{subequations}

\noindent In a two dimensional material, each of these coefficients is actually a 2x2 matrix; the diagonal entries give the response in the direction of an applied field, while the off-diagonal entries give the response in a perpendicular direction (e.g., the Hall conductivity).  We will specifically focus on the longitudinal response in the interchain direction.

The $L^{(il)}$ coefficients in the transport coefficient formulas are defined by\cite[eqs. 3.487, 3.488]{Mahan3}
\begin{subequations}
\begin{align}
J & = -\frac{1}{T}L^{(11)}\nabla(eV) + L^{(12)}\nabla\left(\frac{1}{T}\right)\\
J_E & = -\frac{1}{T}L^{(21)}\nabla(eV) + L^{(22)}\nabla\left(\frac{1}{T}\right)
\end{align}
\end{subequations}
where $J$ is the particle current, or electrical current divided by the charge per particle, and $J_E$ is the energy current.  Note that $L^{(12)}=L^{(21)}$.  In practice, we find the $L^{(il)}$ coefficients in terms of current-current correlation functions as\footnote{Equation \eqref{eq:L_Kubo_formula} is a corrected version of (3.518) from reference \onlinecite{Mahan3}; see the Supplemental Material\cite{SuppMat} for details.}
\begin{align}
L^{(il)} & = \lim_{\omega\rightarrow 0} \lim_{\d\rightarrow 0} \frac{1}{\omega}\left[\frac{-i}{\Omega\b}\int_0^\b d\tau e^{i\omega_n \tau} \langle T_\tau j_l(\tau)j_i(0)\rangle\right]
\label{eq:L_Kubo_formula}\\
&\,\,\,\,\,\,\,\,\,\,\,\,\,\,\,\,\,\,\,\,\,\,\,\,\,\,\,\,\,\,\,\,\,\,\,\,\,\,\,\,\,\,\,\,\,\,\,\,\,\,\,\,\,\,\,\,\,\,\,\,\,\,\,\,\,\,\,\,\,\,\,\,\,\,\,\,\,\,\,\,\,\,\,\,\,\,\,\,\,\,\,\,\,\,\,\,\,\,\,\,\,\,\,\, ^{i\omega_n\rightarrow \omega+i\d}\nonumber
\end{align}
where $j_1$ is the particle current operator $J$ and $j_2$ is the energy current operator $J_E$.  Both are the current operators for the interchain direction.  $\Omega$ is the volume of the system.  Because we calculate the transport coefficients at finite temperature, we perform the calculation using the Matsubara formalism.  $\tau$ is the imaginary time, $\omega_n=2\pi n/\b$ for $n=0,1,2,\cdots$ are the discrete (bosonic) Matsubara frequencies, and $i\omega_n\rightarrow \omega+i\d$ indicates analytic continuation from the positive imaginary axis to just above the positive real axis.  In practice we will take only the real part of $L^{(il)}$, since we are interested specifically in transport.

The current operators we find using \cite{Mukerjee2005}
\begin{subequations}
\begin{align}
J & = \lim_{k\rightarrow 0} \frac{1}{k} \sum_j [N_j,H]e^{ika_c j}\\
J_E & = \lim_{k\rightarrow 0} \frac{1}{k} \sum_j [H_j,H]e^{ika_c j}\label{eq:def_JE_operator}
\end{align}\label{eq:def_J_operators}
\end{subequations}

\noindent in units where $\hbar=1$.  Here $a_c$ is the distance between 1D chains and $N_j$ is the total number operator on chain $j$, $N_j = \sum_k c^\dg_{jk}c_{jk}$.  $H_j$ is the part of the Hamiltonian associated with chain $j$, which includes both the on-chain portion
\begin{subequations}
\begin{equation}
h_j = \sum_{k} E_k c^\dg_{jk} c_{jk}
\end{equation}
and the hopping portion
\begin{equation}
h'_j = - \frac{1}{2}\sum_{kk'} t_{kk'}\left(c^\dg_{j,k} c_{j+1,k'} + c^\dg_{j-1,k} c_{j,k'} \right) +  \text{h.c.}
\end{equation}
\end{subequations}
This leads, after some algebra, to the expressions
\begin{subequations}
\begin{align}
J & = ia_c\sum_{jkk'}t_{kk'}c^\dg_{j-1,k}c_{j,k'} - t_{kk'}^\ast c^\dg_{j,k'}c_{j-1,k}\label{eq:nonint_J_e}\\
J_E & = ia_c\sum_{jkk'} \left[\left(\frac{E_k+E_{k'}}{2}\right) \left(t_{kk'}c^\dg_{j-1,k} c_{jk'}-\text{h.c.}\right)\right]\label{eq:nonint_J_Q}
\end{align}
\end{subequations}
From these current operators and equation \eqref{eq:L_Kubo_formula}, we derive (see Appendix \ref{appendix:nonint_Kubo}) the expression
\begin{equation}
\text{Re}\left[L^{(il)}\right] = \frac{A a_c t^2 v \b^{-n_{il}}}{2\pi^4}\int \frac{g^2(y/\b)y^{n_{il}}}{\left(1+e^y\right)\left(1+e^{-y}\right)}\,\,dy\label{eq:nonint_L}
\end{equation}
where $n_{il}=i+l-2$ (e.g., 0 for $L^{(11)}$), $v$ is the (possibly renormalized by interactions) Fermi velocity, $A$ is a dimensionless number, $t$ is the peak value of the interchain hopping $t=t_{kk}$, $\b$ as usual is $1/T$ (we use units of $k_B=1$), and $g(E)$ is the electronic density of states.  The integral over the dimensionless variable $y=\b E$ runs from $-\infty$ to $\infty$.

The form of the integrand can be intuitively understood from a semiclassical perspective.  If a particle is hopping from an orbital at energy $E$ on one chain to an orbital at energy $E$ on another, then the number of ways that can happen is the number of orbitals at that energy on the first chain, $g(E)$, multiplied by the fraction that are occupied, $(1+e^{\b E})^{-1}$, times the number of orbitals at that energy on the second chain, $g(E)$, multiplied by the fraction that are unoccupied, $(1+e^{-\b E})^{-1}$.  Multiplying all of these factors and integrating over the energy gives
\begin{equation}
\int \frac{g^2(E)}{(1+e^{\b E})(1+e^{-\b E})} dE.\label{eq:semiclassical_intuition}
\end{equation}
This should be proportional to the hopping rate, and therefore to the electrical conductivity.  Indeed, equation \eqref{eq:semiclassical_intuition} looks just like the integrand in equation \eqref{eq:nonint_L} for $L^{(11)}$, which is proportional to the electrical conductivity.  The fact that a semiclassical picture is helpful in understanding equation \eqref{eq:nonint_L} is not too surprising given that our weak hopping assumption is only valid when the temperature is high enough for the interchain hopping to be incoherent.

This is the point in the calculation where the fact that each 1D chain is a Luttinger liquid becomes important.  The density of states for a Luttinger liquid is given by Eq. (61) of reference \onlinecite{Braunecker} as
\begin{equation}
g_{LL}(E)=2\frac{|E/W|^{\g-1}}{2\pi v \Gamma(\g)},\label{eq:LL_dos}
\end{equation}
valid for $E\ll W$, where $W=v/a$ is proportional to the Fermi energy ($E_F\propto k_F^2/m = (k_F/m)/k_F^{-1}\propto v/a$) or bandwidth of the underlying 1D model and $\g$ is a measure of interaction strength in the Luttinger liquid defined by
\begin{equation}
\g=\frac{K+K^{-1}}{2}.\label{eq:def_gamma}
\end{equation}
$K$ is the usual Luttinger liquid interaction parameter, as defined for the Luttinger liquid Hamiltonian below (equation \ref{eq:LL_H}).  (Note that using $K$ for this parameter is a relatively standard convention, used for instance in the book by Giamarchi\cite{GiamarchiBook}, though some authors refer to it as $g$ or $K^2$.\cite{KaneFisher1992,KaneFisher1992b,KaneFisher1996,finiteLLpaper})  $K=1$ corresponds to noninteracting electrons, while $K<1$ corresponds to repulsive interactions and $K>1$ corresponds to attractive interactions.  We have introduced the new parameter $\g$, which is symmetric in $K$ and $K^{-1}$ and thus is independent of whether the interactions happen to be attractive or repulsive.  It always satisfies $\g \geq 1$, and $\g=1$ if and only if the system is noninteracting.

Substituting equation \eqref{eq:LL_dos} into equation \eqref{eq:nonint_L} and using that result in equations \eqref{eq:coeffs_from_L}, we find the following results for the transport coefficients:
\begin{subequations}
\begin{align}
\sigma & = \frac{a_c e^2 t^2}{v T}\left(\frac{T}{W}\right)^{2(\g-1)}\!\!\!\!\!\!\!\!\!\!\times\frac{A}{2\pi^6\Gamma(\g)^2}\int_{y}\frac{|y|^{2(\g-1)}}{\left(1+e^y\right)\left(1+e^{-y}\right)}dy\label{eq:nonint_sigma_LLg}\\
\kappa & = \frac{a_c t^2}{v}\left(\frac{T}{W}\right)^{2(\g-1)}\!\!\!\!\!\!\!\!\times\frac{A}{2\pi^6\Gamma(\g)^2}\int_{y}\frac{y^2|y|^{2(\g-1)}}{\left(1+e^y\right)\left(1+e^{-y}\right)}dy\label{eq:nonint_kappa_LLg}\\
S & = 0
\end{align}
\label{eq:nonint_LLg_results}
\end{subequations}

\noindent Both the thermopower and the second term of equation \eqref{eq:kappa_formula} for the thermal conductivity vanish because $L^{(12)}$ is 0 when the density of states is particle-hole symmetric.  Mathematically this follows because the integrand in equation \eqref{eq:nonint_L} is odd when $g(E)$ is an even function.

\subsection{Correction for nonzero thermopower}

To model a real material and get nonzero thermopower, we can introduce an asymmetry in the band structure.  In particular, the Tomonaga-Luttinger model begins by linearizing a typical 1D band structure around the Fermi points, so we adopt the picture that the Luttinger liquid arises from adding interactions to a 1D electron gas with a typical dispersion $E=\frac{\hbar^2k^2}{2m}\propto k^2$.  In that case, the density of states is $dk/dE\propto E^{-1/2}$.  In our calculations above we have set the Fermi level to $E=0$, in which case the noninteracting density of states becomes
\begin{equation}
g_\text{1D}(E)\propto (E_F+E)^{-1/2}.
\end{equation} 
The Fermi energy is proportional to $v/a$, so for consistency with equation \eqref{eq:LL_dos} we can write it as $E_F = b \, W$ for a dimensionless constant $b$.  Using this 1D density of states as a correction to the Luttinger liquid one gives
\begin{equation}
g(E) = \frac{g_\text{LL}(E)}{\sqrt{1+E/(b \, W)}}\approx g_\text{LL}(E)\left(1-\frac{1}{2}\frac{E}{b\,W}\right).\label{eq:modified_DOS}
\end{equation}
This density of states is a phenomenological way of capturing the real physical behavior of the system which should be accurate enough to find how the thermopower depends on temperature.  The most important features are the violation of particle-hole symmetry by the introduction of a bandwidth and the preservation of the density of states to lowest order in $E/W$ when $E$ is small (near the Fermi energy).  

If we calculate $L^{(12)}$ with equation \eqref{eq:modified_DOS} replacing equation \eqref{eq:LL_dos} as the density of states, we find for the thermopower
\begin{equation}
S = \frac{k_B^2 T}{W e} \times\frac{\int\frac{y^2|y|^{2(\g-1)}}{\left(1+e^{y}\right)\left(1+e^{-y}\right)}dy}{b\int\frac{|y|^{2(\g-1)}}{\left(1+e^{y}\right)\left(1+e^{-y}\right)}dy}
\end{equation}
where Boltzmann's constant has been restored to get the correct final units.  

Note that in principle we could also use the same correction for the conductivities, equations \eqref{eq:nonint_LLg_results}, but any additional terms would be higher order in $k_B T/W$ than those given above.  $k_B T/W$ must be small, otherwise the Tomonaga-Luttinger model, which is based on a linearized band structure (i.e., $W\rightarrow \infty$), would not be applicable at that temperature.

\subsection{Lorenz number}

\begin{figure}
\centering
\includegraphics[width=\columnwidth]{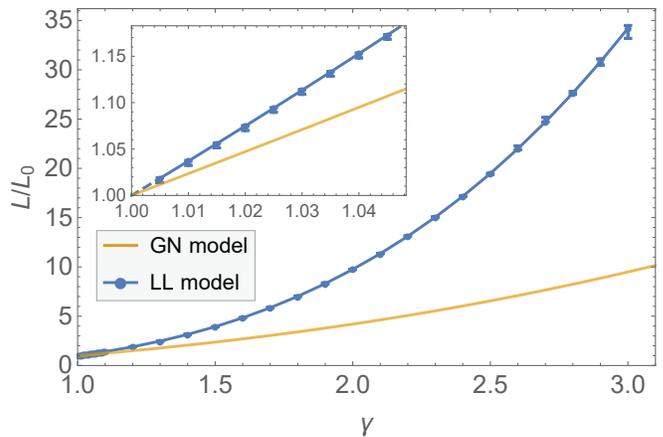}
\caption{(Color online) Lorenz number, $L$, as calculated in the generalized noninteracting (GN) and Luttinger liquid (LL) models.  The Lorenz number is plotted as a function of the interaction strength $\g$ in units of $L_0$, the value expected from the Wiedemann-Franz law.  For both models, we find that $L=L_0$ in the noninteracting case $\g=1$.  Electron-electron interactions ($\g>1$) lead to a violation of the Wiedemann-Franz law; the violation is stronger in the LL model than in the GN model.  The Lorenz number is evaluated at discrete points in the LL model; error bars indicate the precision of numerical results as described in the text.  Lines connecting the data points for the LL model show linear interpolation between adjacent points, and the dashed line below $\g=1.005$ in the inset shows extrapolation to $\g=1$.}
\label{fig:combined_Lorenz_number}
\end{figure}

The expressions for the conductivities, equations \eqref{eq:nonint_LLg_results}, are clear and understandable, but they do contain material-dependent parameters like $a_c$, $v$, and $W$.  To find a robust result that can be tested experimentally, we would like a quantity in which these material-dependent quantities do not appear.  One such parameter is the Lorenz number, 
\begin{equation}
L = \frac{\kappa}{\s T}.
\end{equation}
This is a particularly useful quantity to consider, since the Wiedemann-Franz law states that for a noninteracting system or for a Fermi liquid, the Lorenz number should take a specific value, namely
\begin{equation}
L_0 = \frac{\pi^2}{3}\left(\frac{k_B}{e}\right)^2.
\end{equation}
The Lorenz number for our model can be found by dividing the results from equations \eqref{eq:nonint_LLg_results} to get
\begin{equation}
L = \frac{k_B^2}{e^2} \frac{\int\frac{y^2|y|^{2(\g-1)}}{\left(1+e^{y}\right)\left(1+e^{-y}\right)}dy}{\int\frac{|y|^{2(\g-1)}}{\left(1+e^{y}\right)\left(1+e^{-y}\right)}dy}.\label{eq:nonint_L_LLg}
\end{equation}
As expected from the Wiedemann-Franz law, in the noninteracting limit of $\g=1$ we get precisely $L_0$.  At $\g>1$, this expression for $L$ can be evaluated via numerical integration.  With interactions, $\g>1$, we find that $L>L_0$, violating the Wiedemann-Franz law.  The Lorenz number is plotted as a function of the interaction strength $\g$ in the lower curve in figure \ref{fig:combined_Lorenz_number}.

\begin{figure}
\centering
\includegraphics[width = \columnwidth]{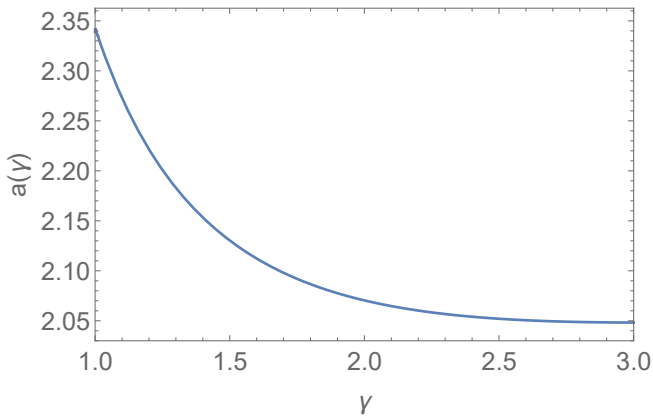}
\caption{The Lorenz number scales as $L/L_0=\g^{a(\g)}$.  For the generalized noninteracting model we find $2<a(\g)<2.35$ for all $\g$, with $L\approx L_0\g^2$ for large $\g$.}
\label{fig:nonint_L_exponent}
\end{figure}

The Lorenz number should scale approximately as $\g^2$ in this model, since the extra two powers of $y$ in equation \eqref{eq:nonint_L} that appear for $L^{(22)}$ (and therefore $\kappa$) but not for $L^{(11)}$ (and therefore $\sigma$) become derivatives with respect to $x$ if the expression is rewritten via Fourier transform; these derivatives act on the Green's function that looks roughly like $f(x)^{-\g}$ and thus pull down two factors of $\g$.  To test that it is indeed the case that $L\approx L_0 \g^2$, we define $a(\g)$ by $L=L_0\g^{a(\g)}$ in which case\begin{equation}
a(\g)=\frac{\log(L/L_0)}{\log(\g)}.\label{eq:def_a_gamma}
\end{equation}
This quantity is plotted in figure \ref{fig:nonint_L_exponent}.  From the plot we see that the exponent $a$ is between $2.35$ and $2$ for all interaction strengths $\g$.  For large $\g$, the scaling of the Lorenz number is close to $\g^2$; for small $\g$, expanding around $\g=1$ gives
\begin{align}
a(\g\approx 1) & = 1-2\log(\pi)+\frac{6}{\pi^2}\!\left(\g_1{\kern 0.001em}'\left(\frac{1}{2}\right)-\g_1{\kern 0.001em}'(1)\!\right)\nonumber\\
&\approx 2.3432
\end{align}
where $\g_1(\nu)$ is a generalized Stieltjes constant\cite{Blagouchine2014}\textsuperscript{,}\footnote{For purposes of calculation, the generalized Stieltjes constant is implemented in the commercial software Wolfram Mathematica as $\g_n(\nu)=\text{StieltjesGamma}[n,\nu]$\cite{WolframSGDef}}.

\subsection{Summary of generalized noninteracting model}

Our most robust predictions are those that do not depend on any material-dependent parameter but the interaction strength.  These are (a) the power law dependencies of $\sigma$, $\kappa$, and $S$ on temperature and (b) the Lorenz number.  We find that 
\begin{subequations}  
\begin{align}
\sigma & \propto T^{2\g-3}\label{eq:nonint_sigma_T_dep}\\
\kappa & \propto T^{2\g-2}\label{eq:nonint_kappa_T_dep}\\
S & \propto T
\end{align}\label{eq:nonint_T_dependence}
\end{subequations}
and
\begin{align}
L & = \frac{k_B^2}{e^2} \frac{\int\frac{y^2|y|^{2(\g-1)}}{\left(1+e^{y}\right)\left(1+e^{-y}\right)}dy}{\int\frac{|y|^{2(\g-1)}}{\left(1+e^{y}\right)\left(1+e^{-y}\right)}dy}\nonumber\\
& \approx L_0 \g^2
\end{align}
In the noninteracting case, $\g=1$, the Lorenz number agrees with the usual Wiedemann-Franz Law.  With either attractive or repulsive interactions, the Wiedemann-Franz law is violated as shown in figure \ref{fig:combined_Lorenz_number}.

%=======================================================================
%=======================================================================

\section{Luttinger liquid model\label{section:LL}}

In the second version of our model, we introduce Luttinger liquid physics much earlier in the analysis.  To do so, we replace assumptions \eqref{assumption:noninteracting} and \eqref{assumption:hopping} with two new, corresponding assumptions:
\begin{enumerate}[(1')]
\setcounter{enumi}{3}%{\ref{assumption:final}}
\item \label{assumption:LL_interacting}Each individual 1D chain is described by the Luttinger liquid Hamiltonian\cite{GiamarchiBook},
\begin{equation}
H = \frac{1}{2\pi}\int dx \left[vK(\nabla\t)^2 + \frac{v}{K}(\nabla\phi)^2\right]\label{eq:LL_H}
\end{equation}
where again we have set $\hbar=1$.  As above, $K$ is a parameter that measures interaction strength and $v$ is the renormalized Fermi velocity. $\phi$ and $\theta$ are bosonic field operators related to the fermion operators by\cite{GiamarchiBook}
\begin{subequations}
\begin{align}
\psi_\a(x) & = U_\a \lim_{a\rightarrow 0} \frac{1}{\sqrt{2\pi a}}e^{i\a k_F x}e^{-i(\a\phi(x)-\t(x))}\\
\psi^\dg_\a(x) & = U_\a^\dg \lim_{a\rightarrow 0} \frac{1}{\sqrt{2\pi a}}e^{-i\a k_F x}e^{i(\a\phi(x)-\t(x))}
\end{align}\label{eq:LL_psi_from_pt}
\end{subequations}

\noindent where $\a$ can be $R$ or $L$ (labeling right-movers versus left-movers) when used as an index and 1 or $-1$, respectively, when used as a multiplicative factor.  The $U_r$ operators are called Klein factors, and are included to make sure that the fermion operators anticommute and that they do not conserve particle number.
\item \label{assumption:LL_hopping}Electrons hop between real-space localized orbitals.  The hopping strength is sharply peaked in $|x-x'|$, where $x$ and $x'$ are the locations along the two chains, measured from the same ``center'' point (so that all the ``$x=0$'' points lie on a line perpendicular to the chains).  In the thermodynamic limit, a delta-function hopping in real space is consistent with the sharply peaked hopping in Fourier space from assumption \eqref{assumption:hopping} from the first version of our model (see Appendix \ref{appendix:correspondence}).  We also assume that right-movers on one chain can only hop to right-movers on the adjacent chain and the same for left-movers; this is needed for consistency with the approximate momentum conserving hopping in the generalized noninteracting model.
\label{assumption:LL_final}
\end{enumerate}

\noindent Including both on-chain and hopping terms, the Hamiltonian for this second version of our model is:
\begin{widetext}
%\begin{align}
%H & = \sum_j \, H_j = \sum_j \, h_j + h'_j\nonumber\\
%h_j & = \frac{1}{2\pi}\int \! dx \left[vK\left(\nabla \t_j\right)^2 + \frac{v}{K}\left(\nabla \phi_j\right)^2\right]\label{eq:LL_model_H}\\
%h'_j & =\! -\frac{1}{2}\!\sum_{\a\b}\!\int\! dx\,dx'\!\left[\!\begin{array}{l}t_{\a\b}(x-x')\left(\!\!\!\begin{array}{c}\psi_{j\a}^\dg(x)\psi_{j+1,\b}(x')\\ +\, \psi_{j-1,\a}^\dg(x)\psi_{j\b}(x')\end{array}\!\!\!\right)\\ + \,\, \text{h.c.}\end{array}\!\!\!\right]\nonumber
%\end{align}
\begin{align}
H & = \sum_j \, H_j = \sum_j \, h_j + h'_j\nonumber\\
h_j & = \frac{1}{2\pi}\int \! dx \left[vK\left(\nabla \t_j\right)^2 + \frac{v}{K}\left(\nabla \phi_j\right)^2\right]\label{eq:LL_model_H}\\
h'_j & =\! -\frac{1}{2}\!\sum_{\a\b}\!\int\! dx\,dx'\!\left[t_{\a\b}(x-x')\left(\psi_{j\a}^\dg(x)\psi_{j+1,\b}(x')+\, \psi_{j-1,\a}^\dg(x)\psi_{j\b}(x')\right) + \,\, \text{h.c.}\right]\nonumber
\end{align}

\subsection{Calculation of transport coefficients}

\noindent As in the generalized noninteracting model, to find the transport coefficients we first find operators for the electrical and energy currents.  This can be done using equations \eqref{eq:def_J_operators} just as before, but with the new definitions for $h_j$ and $h'_j$.  The results (for some details of the calculation, see Appendix \ref{appendix:LL}) are
%\begin{widetext}
\begin{subequations}
\begin{align}
J & = -ia_c\sum_j\sum_{\a\b = R,L} \int t_{\a\b}(x-x')\left[\psi_{j\a}^\dg(x)\psi_{j-1,\b}(x') - \psi_{j-1,\a}^\dg(x)\psi_{j\b}(x')\right]\,dx\,dx'\label{eq:LL_J_e}\\
J_E & = -\frac{ia_c v}{2}\sum_{j\a\b}\int t_{\a\b}(x-x')\left[\left([\nabla_j]^\a_{x} +[\nabla_{j-1}]^\b_{x'}\right)\psi^\dg_{j\a}(x)\psi_{j-1,\b}(x') - \left([\nabla_{j-1}]^\a_{x} +[\nabla_j]^\b_{x'}\right)\psi^\dg_{j-1,\a}(x)\psi_{j\b}(x')\right]\,dx\,dx'
\label{eq:LL_J_E}
\end{align}
\end{subequations}
%\end{widetext}
where
\begin{equation}
[\nabla_j]^\a_y = \a K\nabla\t_j(y) - K^{-1}\nabla\phi_j(y).\label{eq:def_nabla}
\end{equation}

\noindent Unlike in the generalized noninteracting model, we do not find a single simple formula like equation \eqref{eq:nonint_L} that gives all the transport coefficients.  Instead, the particularly nice expressions that we find are for the current-current correlators in terms of the Green's function for a single Luttinger liquid:
%\begin{widetext}
\begin{subequations}
\begin{align}
\langle J(\tau) J\rangle & = -2N_c L\left(\frac{a_c t}{2\pi}\right)^2\sum_\a\int dx\, G_\a(x,\tau)G_\a(-x,-\tau)\label{eq:LL_JeJeCorrelator_unspecified_Greens}\\
\langle J_E(\tau) J_E\rangle & = -2N_cL\g^2\left(\frac{a_c v t}{2\pi}\right)^2\sum_\a\int dx\,\big[(k_F+i\a\partial_x)G_\a(x,\tau)\big]\times\big[(k_F-i\a\partial_x)G_\a(-x,-\tau)\big]\label{eq:LL_JEJECorrelator_unspecified_Greens}\\
\langle J_E(\tau) J\rangle & = 2v\g N_c L\left(\frac{a_c t}{2\pi}\right)^2 \sum_{\a}\int dx\,
G_\a(x,\tau)(k_F-i\a\partial_x) G_\a(-x,-\tau)\label{eq:LL_JEJeCorrelator_unspecified_Greensv2}
\end{align}\label{eq:JJCorrelators_unspecified_Greens}
\end{subequations}

\noindent For the Green's function we use the expression\cite{finiteLLpaper,Braunecker}
\begin{equation}
G_\a(x,\tau) = -\frac{e^{i\a k_F x}}{2\pi a}\left[\frac{-ia}{\frac{v\b}{\pi}\sinh\left(\frac{x-iv\tau}{v\b/\pi}\right)}\right]^{\frac{\g-\a}{2}}\left[\frac{ia}{\frac{v\b}{\pi}\sinh\left(\frac{x+iv\tau}{v\b/\pi}\right)}\right]^{\frac{\g+\a}{2}}\label{eq:LL_Greens}
\end{equation}
and we are then able to perform the integration over $x$ exactly, getting results in terms of the Appell hypergeometric function $F_1$ as defined in \S 16.13 of reference \onlinecite{NIST:DLMF}.\footnote{For purposes of calculation, the function $F_1$ is implemented in the commercial software Wolfram Mathematica as $F_1(a;b_1,b_2;c;x,y)=\text{AppellF1}[a,b1,b2,c,x,y]$\cite{WolframAppellDef}}  As an example, the result for $\langle J(\tau)J\rangle$ is
\begin{equation}
\langle J(\tau') J\rangle= 4N_c L\left(\frac{a_c t}{2\pi}\right)^2\frac{2a}{(2\pi a)^2}\left(\frac{2\pi a}{v\b}\right)^{2\g-1}\bigg(2 f(\g,\tau',1,1)-\cos(2\tau')\big(f(\g,\tau',0,1) + f(\g,\tau',2,1)\big)\bigg)
\label{eq:LL_JeJeCorrelatorFull}
\end{equation}
where $\tau'$ is a scaled version of the imaginary time, $\tau' = \tau\pi/\b$, and 
\begin{equation}
f(\g,\tau,n,m) = \frac{F_1(\g+n;\g+m,\g+m;\g+n+1;e^{2i\tau},e^{-2i\tau})}{\g+n}.\label{eq:f_from_F1}
\end{equation}
\end{widetext} 
The analogous expressions for the other two current-current correlators are longer and more complex, and thus proportionally less enlightening.  We present them in Appendix \ref{appendix:LL} for the edification of the interested reader.

The next step is to evaluate each of the $L^{(il)}$ coefficients using equation \eqref{eq:L_Kubo_formula}.  In the previous model it was possible to perform the Fourier transform and analytic continuation analytically, but here we must perform the $\tau$ integrals of the current-current correlators numerically for each Matsubara frequency and then numerically perform the analytic continuation and limits.  The procedure we follow is discussed further in Appendix \ref{appendix:LL}, and in great detail in the Supplemental Material\cite{SuppMat}.

For each transport coefficient, we get a numerical part from the procedure mentioned above and a prefactor that contains all the dimensionful quantities, notably the dependence on temperature.  Including for now only the dimensionful quantities, we find
%\begin{subequations}
%\begin{align}
%\sigma & \propto \frac{a_c}{T^2}\left(\frac{qt}{\hbar}\right)^2\frac{1}{a}\left(\frac{a}{v\b}\right)^{2\g-1}\\
%\kappa & \propto \g^2 \frac{a_c}{T^3}\left(\frac{v t}{\hbar}\right)^2\frac{1}{a^3}\left(\frac{a}{v\b}\right)^{2\g+1}\\
%S & = 0
%\end{align}
%\end{subequations}
\begin{subequations}
\begin{align}
\sigma & \propto \frac{a_c a \,q^2 t^2}{v^2}\left(\frac{a}{v\b}\right)^{2\g-3}\\
\kappa & \propto \frac{a_c t^2}{v}\left(\frac{a}{v\b}\right)^{2\g-2}\\
S & = 0
\end{align}\label{eq:LL_transport_dimensions}
\end{subequations}

\noindent Recalling that the energy scale $W$ introduced in the Luttinger liquid density of states, equation \eqref{eq:LL_dos}, was $W=v/a$, the dependence of the transport coefficients on the material-dependent parameters $a_c$, $a$, and $v$ in this model (equations \ref{eq:LL_transport_dimensions}) precisely matches what we found in the generalized noninteracting model (equations \ref{eq:nonint_LLg_results}).

In the generalized noninteracting model, we introduced a correction to the density of states to find a nonzero thermopower.  Due to the complexity of the full Luttinger liquid model, we consider the equivalent correction here to be beyond the scope of this paper. 

\subsection{Lorenz number}
The numerical analytic continuation has not yet been needed for the results presented above.  We would like, however, to evaluate the Lorenz number numerically as a function of the interaction strength, $\g$, just as in the generalized noninteracting model.  For that calculation, the full numerics are needed.  

To compute the precise transport coefficients, for each interaction strength $\g$ we must separately evaluate the Fourier transform of the current-current correlation functions at a number of Matsubara frequencies, fit an analytic function to these results, analytically continue the function, and then take the limits as the frequency $\omega$ and the infinitesimal parameter $\d$ go to 0.  (For details, see Appendix \ref{appendix:LL}.)  

Due to the complexity of the correlation functions (for instance equation \ref{eq:LL_JeJeCorrelatorFull}), the calculation of each Fourier transform, and thus the calculation of transport coefficients for each interaction strength $\g$, is very computationally expensive.  We therefore evaluate the Lorenz number for a limited number of values of the interaction strength, with a higher density around $\g=1$ to make sure that the results in the noninteracting limit are reliable.  The results are shown for $\g$ in the range 1 to 3 by the discrete data points in figure \ref{fig:combined_Lorenz_number} (connected by linear interpolation for visual clarity).  An inset shows a detail of $\g\in[1,1.05]$; from the inset it is clear that in the noninteracting limit the Lorenz number approaches the expected value from the Wiedemann-Franz law.

The error bars on the Luttinger liquid model data in figure \ref{fig:combined_Lorenz_number} indicate the numerical precision of the Lorenz number for each $\g$.  We compute the numerical integral for each Fourier transform with a relative precision of $10^{-10}$, and allowing the values of the Fourier transforms to vary within this range and recomputing the Lorenz number gives a sharply peaked distribution of possible values of $L$.  The error bars in the figure show one standard deviation of this distribution for each interaction strength $\g$.

Comparing the results of the full Luttinger liquid model with the corresponding results for the generalized noninteracting model, as shown in the upper and lower curves respectively in figure \ref{fig:combined_Lorenz_number}, we see that the full Luttinger liquid model exhibits a stronger violation of the Wiedemann-Franz law with increasing interaction strength.  We argued that in the generalized noninteracting model the Lorenz number should scale as $\g^2$ because the two extra factors of energy for $L^{(22)}$ relative to $L^{(11)}$ in equation \eqref{eq:nonint_L} act, in a real-space representation, as derivatives of the Green's function.  For the full Luttinger liquid model, we can make a similar argument that $L/L_0\approx \g^4$.  There are indeed two derivatives acting on the Green's function in the expression for $\langle J_E(\tau)J_E\rangle$, equation \eqref{eq:LL_JEJECorrelator_unspecified_Greens}, that are not present in the expression for $\langle J(\tau)J\rangle$, equation \eqref{eq:LL_JeJeCorrelator_unspecified_Greens}, giving rise to the same two factors of $\g$ as in the generalized noninteracting model.  

There are additionally two factors of $\g$ in the prefactor in the expression for $\langle J_E(\tau)J_E\rangle$, which come from the $[\nabla_j]^\a_x$ operators in the expression for the energy current operator, equation \eqref{eq:LL_J_E}, and are thus missing from the generalized noninteracting model because there the energy current operator was derived in the noninteracting limit where $\g=1$.  With these two additional factors of $\g$ included, we find that the Lorenz number should scale approximately as $L/L_0\approx \g^4$.

\begin{figure}
\centering
\includegraphics[width=\columnwidth]{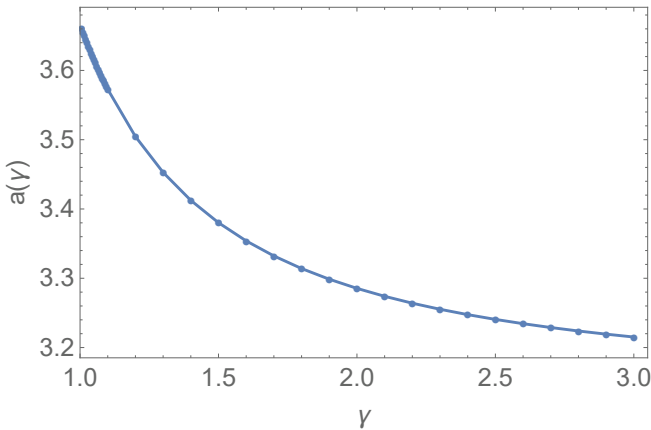}
\caption{\label{fig:LL_L_exponent}Exponent $a$ in $L/L_0=\g^{a(\g)}$ for the Luttinger liquid model.  The dependence on $\g$ is stronger than in the generalized noninteracting model. Lines are given by linear interpolation between adjacent data points, and error bars are omitted for clarity.}
\end{figure}

This argument neglects the full complexity of the correlation functions, so to find more precisely how the Lorenz number scales with $\g$ we again introduce the function $a(\g)$ defined by equation \eqref{eq:def_a_gamma}, $L/L_0 = \g^{a(\g)}$.  This is plotted in figure \ref{fig:LL_L_exponent}.  We find that $L/L_0$ satisfies $\g^{3.2}<L/L_0<\g^{3.7}$ for $\g\leq 3$.  This is a slightly weaker dependence than the predicted $\g^4$, but it is still a much stronger violation of the Wiedemann-Franz law than $L/L_0\approx \g^2$ from the generalized noninteracting model.

\subsection{Summary of Luttinger liquid model}

As in the generalized noninteracting model, it is useful to summarize those results that do not depend on any material-dependent parameter apart from the interaction strength.  For the dependence of the conductivities on temperature, we find the same power laws as in the generalized noninteracting model, namely $\sigma \propto T^{2\g-3}$ and $\kappa \propto T^{2\g-2}$.  For the Lorenz number we find a stronger violation of the Wiedemann-Franz law than in the generalized noninteracting model.  We analytically estimate that
\begin{equation}
L \approx L_0 \g^4
\end{equation}
and numerically observe that 
\begin{equation}
L_0\g^{3.2}<L<L_0\g^{3.7}
\end{equation}
The precise dependence of the Lorenz number on the interaction strength is shown in figure \ref{fig:combined_Lorenz_number}.  In the noninteracting case, $\g=1$, the Lorenz number is $L_0$, the expected value from the Wiedemann-Franz Law.

%=======================================================================
%=======================================================================

\section{Discussion and analysis}\label{section:discussion}

We have analyzed two different models for weakly coupled Luttinger liquids, finding in both cases the electrical and thermal conductivity.  In terms of the interaction parameter $\g$, the conductivities scale in both models as $\sigma \propto T^{2\g-3}$ and $\kappa \propto T^{2\g-2}$.  In both cases we find a violation of the Wiedemann-Franz law with increasing interaction strength; for the generalized noninteracting model $L\approx L_0\g^2$ as shown in figures \ref{fig:combined_Lorenz_number} and \ref{fig:nonint_L_exponent}, while for the Luttinger liquid model $L\approx L_0\g^{3.2}$ as shown in figures \ref{fig:combined_Lorenz_number} and \ref{fig:LL_L_exponent}.  This type of violation of the Wiedemann-Franz law as a power of the interaction strength is similar to the result of Kane and Fisher\cite{KaneFisher1996}, although the precise dependence is of course different since our models describe a different physical system.  In the generalized noninteracting model we also find a nonzero expression for the thermopower if we correct the density of states to account for particle-hole symmetry breaking, in which case $S\propto T$.  This linear dependence of thermopower on temperature, which matches the expected behavior in a Fermi liquid, was also found by Kane and Fisher in their coupled chain model\cite{KaneFisher1996}.

The violation of the Wiedemann-Franz law that we observe in both models is an indication that Luttinger liquid behavior survives when 1D chains are coupled to form a two-dimensional material.  Just how large is the violation in practice?  Experimental measurements\cite{Vescoli1998} and theoretical calculations\cite{Schulz1990,Ejima2005,Ejima2006,Sirker2012,Affleck2013} have found Luttinger parameters in a typical range of about $0.2$ through $1.5$, corresponding to values of $\g$ up to about 3.  In both our models, $\g=3$ would lead to a large violation of the Wiedemann-Franz law by an order of magnitude or more, an easily measurable effect that could be observed in experiments.

The results summarized here are all independent of any material-dependent parameters apart from the Luttinger liquid interaction parameter, which makes them good candidates for experimental testing and verification on any system with strong anisotropy that might lead to quasi-one-dimensional behavior.  One very direct application of our theory would be to highly anisotropic crystals, as they typically have electron hopping strength along one axis which is at least an order of magnitude stronger than the hopping along the other two axes\cite{Vescoli1998}.  For temperatures between the two hopping scales, it would be reasonable to treat the system as a collection of weakly coupled 1D chains as we have done here, and by the nature of the crystal they form an ordered array, again matching our model.  Such anisotropic crystals are known to show strong violations of the Wiedemann-Franz law, especially in the Hall direction in a magnetic field~\cite{Wakeham2011}.  By comparing the measured violations of the Wiedemann-Franz law in these systems with our predictions, it should be possible to estimate the effective Luttinger parameter $K$ for the constituent one-dimensional chains.  Conversely, if $K$ is independently known then such measurements would serve as a verification of our predictions.

Applying our theory to polymer films, the original motivation of the work, requires some additional work since the films are partially amorphous.  One approach would be to treat the polymer film as a polycrystal, consisting of randomly oriented grains; within each grain, the polymers form an ordered array to which our theory directly applies.  The overall transport properties of the polymer film could then be found by averaging using methods like those discussed in reference \onlinecite{Isichenko1992}.  The precise level of alignment of polymers can also vary significantly between films\cite{Stanford_experiment_paper,Bubnova2013,Steyrleuthner2014}, and more work is needed to properly take this into account.  One experimental result on polymer films which is clearly consistent with our calculations is the fact that some polymers show conductivity increasing with temperature, while others show the opposite behavior\cite{Bubnova2013}.  We find that $\sigma$ increases with temperature if the interaction strength is large enough, $\g>3/2$, but decreases with increasing temperature for $1\leq\g<3/2$.

Numerical studies of transport and other dynamical properties in quasi-one-dimensional systems have made great progress since the advent of matrix product state algorithms for time dependence~\cite{vidalrg,whitetdmrg,schollwoeck}.  In the case of a single chain, it is possible to see the characteristic power laws of Luttinger liquid behavior~\cite{huangkarrasch}, and while coupled chains are considerably more demanding, it has been possible to access at least some excited-state properties~\cite{jameskonik}.  Coupled-chain numerical studies could in principle provide a more precise and tunable ``numerical laboratory'' to test our predictions than current polymer experiments.  

There are a number of ways that our models could be extended for future work.  We have dealt only with spinless Luttinger liquids, so a spin sector could be added.  Due to the spin-charge separation in Luttinger liquids, this would be a relatively simple change and would just result in extra additive contributions to some $L^{(il)}$ coefficients.  The models could also be made more complete via the addition of disorder and by going to higher order in the perturbation theory in the interchain hopping strength.  The latter two corrections would be potentially quite difficult, though disorder could be added at a relatively late stage in the calculation by modifying the density of states as used in equation \eqref{eq:nonint_L} or the Green's function in equations \eqref{eq:JJCorrelators_unspecified_Greens}.  

To implement these or other extensions of our model, if the goal is only to find how transport properties depend on temperature then it will apparently be sufficient to use a noninteracting model for most of the calculation as in section \ref{section:nonint_model}; if the precise values of the transport coefficients are needed, such as for calculating the Lorenz number, then a more complete calculation, as in section \ref{section:LL}, will be required.

%=======================================================================
%=======================================================================

\begin{acknowledgments}
We would like to acknowledge generous support from the Air Force Office of Scientific Research Multidisciplinary Research Program of the University Research Initiative (AFOSR MURI FA9550-12-1-0002), as well as travel support from a Simons Investigatorship.  We also want to thank Christoph Karrasch, Takahiro Morimoto, Snir Gazit, Benjamin Ponedel, and Len Evans for helpful conversations.
\end{acknowledgments}

%=======================================================================
%=======================================================================

\appendix
\section{Details of generalized noninteracting model\label{appendix:nonint_Kubo}}

In the main body of the paper, we focused on the key results of our work and restricted discussion of the calculations to the general formalism that we used.  In this appendix and the ones that follow, we discuss key steps of the calculations, especially those in which we use one of our assumptions.  We also provide some intermediate results such as the current-current correlation functions for the Luttinger liquid model in terms of the hypergeometric function $F_1$.  For a reader interested in seeing more details, we have made our full calculations available in the Supplemental Material\cite{SuppMat}.

\subsection{Current operators}

The computation of the particle and energy current operators, as given in equations \eqref{eq:nonint_J_e} and \eqref{eq:nonint_J_Q}, from equations \eqref{eq:def_J_operators} involves computing the commutators $[N_j,H]$ and $[H_j,H]$ respectively.  In each case, the best way to proceed with the calculation is to break the Hamiltonian into the on-chain and interchain coupling pieces, $H=\sum_i h_i + h'_i$.  As the on-chain Hamiltonian conserves the total number of electrons on the chain, it must commute with the number operator on each chain, so that $[N_j,H]=\sum_i [N_j,h'_i]$.  Similarly, the on-chain Hamiltonians for different chains all commute so that 
\begin{equation}
[H_j,H]=\sum_i [h_j,h'_i]+[h'_j,h_i]+[h'_j,h'_i].
\end{equation}
We also neglect the last term as it contains two powers of the interchain hopping strength and thus is not lowest order in our perturbative calculation.  The remainder of the derivation of the current operators consists of computing the commutators and then observing that half the terms can have their index shifted by 1 in the sum over $j$ from equations \eqref{eq:def_J_operators}, in which case the limit as $k\rightarrow 0$ gives 
\begin{equation}
\frac{1-e^{ik a_c}}{k}\rightarrow -i a_c.\label{eq:index_k_to_0_trick}
\end{equation}
For further details, see the Supplemental Material\cite{SuppMat}.

\subsection{Finding \texorpdfstring{$L^{(il)}$}{L}}

The first step in finding the transport coefficients is to find the time evolution of the current operators.  In imaginary time $\tau = it$, the time evolution is given by
\begin{equation}
J(\tau) = e^{H\tau}J e^{-H\tau}.
\end{equation}
In general this would be a very difficult calculation, but it is made much easier by the fact that we do the calculation only to lowest order in the interchain hopping, which allows us to drop the hopping terms entirely from the Hamiltonian used for the time evolution,
\begin{equation}
H\rightarrow H_0 = \sum_i h_i.
\end{equation}
This means that the time evolution operator acts separately on each creation and annihilation operator in equations \eqref{eq:nonint_J_e} and \eqref{eq:nonint_J_Q}.  The resulting time-dependent current operators are 
\begin{widetext}
\begin{subequations}
\begin{align}
J(\tau) & = ia_c\sum_{jkk'}e^{\tau(E_k-E_{k'})}t_{kk'}c^\dg_{j-1,k}c_{j,k'} - e^{\tau(E_{k'}-E_k)}t_{kk'}^\ast c^\dg_{j,k'}c_{j-1,k}\\
J_E(\tau) & = ia_c\sum_{jkk'} \left[\left(\frac{E_k+E_{k'}}{2}\right) \left(e^{\tau(E_k-E_{k'})}t_{kk'}c^\dg_{j-1,k} c_{j,k'}-e^{\tau(E_{k'}-E_k)}t_{kk'}^\ast c_{jk'}^\dg c_{j-1,k}\right)\right]
\end{align}
\end{subequations}
%\end{widetext}
We then calculate the current-current correlators.  In this appendix we show only the calculations for $\langle J(\tau)J\rangle$, as the others are quite similar.  The brackets $\langle \cdot \rangle$ indicate a thermal expectation value defined as usual by 
\begin{equation}
\langle \mathcal{O}\rangle = \text{Tr}[e^{-\b H}\mathcal{O}]/\text{Tr}[e^{-\b H}] = \text{Tr}[e^{-\b H}\mathcal{O}]/Z
\end{equation}
As with the time evolution, the lowest order result in the interchain hopping can be found by simply dropping the interchain hopping terms from $H$ in the thermal density matrix, $e^{-\b H}\rightarrow e^{-\b H_0}$, in which case the expression for the current-current correlator can be written in terms of expectation values on single chains,
%\begin{widetext}
\begin{equation}
\langle J(\tau)J\rangle = a_c^2 \sum_{jkk'}|t_{kk'}|^2\left(e^{\tau(E_{k'}-E_k)} (1-\langle n_{j-1,k}\rangle)\langle n_{j,k'}\rangle + e^{\tau(E_k-E_{k'})}\langle(1-\langle n_{j,k'}\rangle)\langle n_{j-1,k} \rangle\right),
\end{equation}
where as usual the number operator is given by $n=c^\dg c$.  The expectation value of each number operator is just given by the Fermi-Dirac distribution and is independent of the chain number $j$ so this becomes
\begin{small}
\begin{subequations}
\begin{align}
\langle J(\tau) J\rangle & = N_c a_c^2 \sum_{kk'}|t_{kk'}|^2\left[\frac{e^{\tau(E_{k'}-E_k)}}{\left(1+e^{-\b E_k}\right)\left(1+e^{\b E_{k'}}\right)} + \frac{e^{\tau(E_k-E_{k'})}}{\left(1+e^{\b E_k}\right)\left(1+e^{-\b E_{k'}}\right)}\right]\\
& = N_c a_c^2 \left(\frac{L}{2\pi}\right)^2\int_{kk'}|t(k,k')|^2\left[\frac{e^{\tau(E(k')-E(k))}}{\left(1+e^{-\b E(k)}\right)\left(1+e^{\b E(k')}\right)} + \frac{e^{\tau(E(k)-E(k'))}}{\left(1+e^{\b E(k)}\right)\left(1+e^{-\b E(k')}\right)}\right]dk\,dk'\\
& = 2N_c a_c^2 \left(\frac{L}{2\pi}\right)^2\int_{EE'}|t(E,E')|^2 g(E)g(E')\left[\frac{e^{\tau(E'-E)}}{\left(1+e^{-\b E}\right)\left(1+e^{\b E'}\right)} + \frac{e^{\tau(E-E')}}{\left(1+e^{\b E}\right)\left(1+e^{-\b E'}\right)}\right]dE\,dE'\\
& = 4N_c a_c^2 \left(\frac{L}{2\pi}\right)^2\int_{EE'}|t(E,E')|^2 g(E)g(E')\left[\frac{e^{\tau(E-E')}}{\left(1+e^{\b E}\right)\left(1+e^{-\b E'}\right)}\right]dE\,dE'\label{eq:nonint_JeJe_intermediate}
\end{align}
\end{subequations}
\end{small}
\end{widetext}
where in successive steps we have (1) rewritten the sum over $k$ as an integral over a continuous variable, (2) converted to an integral over energy $E$, with $t(E,E')$ defined by $t(E(k),E(k'))=t(k,k')$ for all $k$ and $k'$, also getting a factor of 2 for the two branches of the dispersion, and (3) recognized that the two terms are the same if, as we assume, $t(E,E')=t(E',E)$.

In the continuum case, the hopping $t(k,k')$ becomes a Dirac delta function.  Thus one factor of $t(E,E')$ collapses the two integrals into one, leaving $t(E,E)\propto \d(0)$.  The appearance of the apparently infinite quantity $\d(0)$ is not a problem because when we do the conversion from a sum over $k$ to an integral, $t_{kk'}$ (which we initially viewed as a sharply peaked, perhaps Gaussian, function) becomes
\begin{equation}
t_{kk'} = t e^{-(k-k')^2L^2/\pi} \rightarrow t(k,k') = \frac{t}{L}\d(k-k')\label{eq:tkk}
\end{equation}
with $\d(0) = L$.  (The precise form of $t_{kk'}$ that we use here is discussed in Appendix \ref{appendix:correspondence} and more thoroughly in the Supplemental Material\cite{SuppMat}.)  This means that $t(E,E)$ is actually just equal to $t$, a constant.  Using this form for $t(E,E')$ gives
\begin{equation}
\langle J(\tau) J\rangle = \frac{4N_cLv\left(a_c t\right)^2}{(2\pi)^2}\!\int \!\!\frac{g^2(E)}{\left(1+e^{\b E}\right)\left(1+e^{-\b E}\right)}dE.\label{eq:nonint_JeJe_corr}
\end{equation}
The corresponding expressions for $\langle J_E(\tau)J_E\rangle$ and $\langle J_E(\tau)J\rangle$ are quite similar, but with extra factors of $E$ in the integrand.  The most noteworthy aspect of this expression from a calculational perspective is that it does not depend on the imaginary time $\tau$ at all.  Then when we calculate the Fourier transform in the equation for $L^{(il)}$, equation \eqref{eq:L_Kubo_formula}, the integral over $\tau$ is just
\begin{equation}
\int_0^\b e^{i\omega_n\tau}d\tau = \b\,\d_{n0}, 
\end{equation}
proportional to a Kronecker delta in the Matsubara frequency.  The analytic continuation of this function is not well-defined, so it is not immediately obvious how to convert the Matsubara correlation function to a retarded one.  This problem, however, arises only when the interaction strength is precisely 0, since otherwise the $\tau$ dependence would not have vanished.  Thus this should be regularized by some small amount of interaction (or by disorder or some other mechanism) in any realistic system.  We thus convert to the dimensionless variable $\tau'=\tau\pi/\b$ and let
\begin{equation}
A=\text{Re}\left(\lim_{n\rightarrow 0}\lim_{\d'\rightarrow 0} \frac{-i}{n} \left[\int_0^\pi e^{2in\tau'} d\tau'\right]_{in\rightarrow n+i\d'}\right).
\end{equation}
This constant corresponds to $F_\a(0)$ in equation (3) of reference \onlinecite{Giamarchi2000_2013}.  Rewriting the expression for $L^{(il)}$ from equation \eqref{eq:L_Kubo_formula} in terms of $\tau'$ and then substituting both the current-current correlator from equation \eqref{eq:nonint_JeJe_corr} (and the corresponding results for $\langle J_E(\tau)J_E\rangle$ and $\langle J_E(\tau)J\rangle$) and the definition of $A$, we get equation \eqref{eq:nonint_L}, our final result for $L^{(il)}$ in the generalized noninteracting model.

%=======================================================================
%=======================================================================

\section{Details of Luttinger liquid model\label{appendix:LL}}

The calculations for the Luttinger liquid model are substantially more complex.  Here we highlight some interesting features particularly of the calculation of the thermal current operator and the correlation function $\langle J_E(\tau) J_E\rangle$.  We also present expressions for $\langle J_E(\tau) J_E\rangle$ and $\langle J_E(\tau) J\rangle$ in terms of the hypergeometric function $F_1$, and we discuss the method we use for numerical analytic continuation to get the transport coefficients from the correlation functions.

\subsection{Thermal current operator}

We calculate the current operators in the full Luttinger liquid model using the same approach as in the generalized noninteracting model.  The additional complication in the calculation comes from the more complete Hamiltonian (equation \ref{eq:LL_model_H}) and in particular from the on-chain part.  As with the calculation of the thermal current operator in the previous model as discussed in Appendix \ref{appendix:nonint_Kubo}, the commutator $[H_j,H]$ from equation \eqref{eq:def_JE_operator} has only two pieces that are neither 0 nor negligible in the atomic limit,
\begin{equation}
[H_j,H]\rightarrow \sum_i [h_j,h'_i] + [h'_j,h_i] = \sum_i [h'_j,h_i]-[h'_i,h_j].
\end{equation}
Terms in the commutator $[h'_i,h_j]$ look like $[\psi^\dg_{i+1,\a}(x)\psi_{i\b}(x'),(\nabla \t_j(\tilde{x}))^2]$.  To compute these kinds of terms, we need the canonical commutation relations between the bosonic field operators $\phi$ and $\theta$, which are given by\cite{GiamarchiBook}:
\begin{subequations}
\begin{align}
[\phi_i(x),\partial_{x'}\t_j(x')] & = i\pi\d_{ij}\d(x'-x)\\
[\phi_i(x),\t_j(x')] & = i\frac{\pi}{2}\d_{ij}\text{sign}(x'-x)\\
[\phi_i(x),\phi_j(x')] & = [\t_i(x),\t_j(x')] = 0
\end{align}\label{eq:LL_boson_commutators}
\end{subequations}

\noindent We then write out the Fermionic operators $\psi$ and $\psi^\dg$ in terms of $\phi$ and $\theta$ using equations \eqref{eq:LL_psi_from_pt} and use the bosonic commutators from equations \eqref{eq:LL_boson_commutators} to show
\begin{subequations}
\begin{align}
[\psi_{i\a}(x),\nabla \t_j(x')] & = \a\pi\d_{ij}\d(x-x')\psi_i(x)\label{eq:LL_JE_ferm_theta_commutator}\\
[\psi^\dg_{i\a}(x),\nabla \t_j(x')] & = -\a\pi\d_{ij}\d(x-x')\psi^\dg_i(x)\label{eq:LL_JE_ferm_dg_theta_commutator}\\
[\psi_{i\a}(x),\nabla \phi_j(x')] & = -\pi\d_{ij}\d(x-x')\psi_i(x)\label{eq:LL_JE_ferm_phi_commutator}\\
[\psi^\dg_{i\a}(x),\nabla \phi_j(x')] & = \pi\d_{ij}\d(x-x')\psi^\dg_i(x)\label{eq:LL_JE_ferm_dg_phi_commutator}
\end{align}\label{eq:LL_JE_ferm_bose_commutators}
\end{subequations}

\noindent Combining these commutators with the rule $[AB,C]=A[B,C]+[A,C]B$, we additionally find that 
\begin{widetext}
\begin{subequations}
\begin{align}
[\psi^\dg_{i\a}(\tilde{x})\psi_{i+1,\b}(x),(\nabla \t_j(x'))^2] & = \left[\begin{array}{c}2\pi\nabla\t_j(x')\left(\b\d(x-x')\d_{i+1,j} - \a\d(x-\tilde{x})\d_{ij}\right) \\+ \pi^2\left(\b\d(x-x')\d_{i+1,j} - \a\d(x-\tilde{x})\d_{ij}\right)^2\end{array}\right]\psi^\dg_{i\a}(\tilde{x})\psi_{i+1,\b}(x)\\
[\psi^\dg_{i\a}(\tilde{x})\psi_{i+1,\b}(x),(\nabla \phi_j(x'))^2] & = \left[\begin{array}{c}-2\pi\nabla\phi_j(x')\left(\d(x-x')\d_{i+1,j} - \d(x-\tilde{x})\d_{ij}\right) \\+ \pi^2\left(\d(x-x')\d_{i+1,j} - \d(x-\tilde{x})\d_{ij}\right)^2\end{array}\right]\psi^\dg_{i\a}(\tilde{x})\psi_{i+1,\b}(x)
\end{align}
\end{subequations}
and hence
\begin{equation}
[\psi^\dg_{i\a}(\tilde{x})\psi_{i+1,\b}(x),h_j] = v\left[\begin{array}{c}\d_{i+1,j}\left(\b K\nabla\t_j(x)-\a K^{-1}\nabla\phi_j(x)\right)-\d_{ij}\left(K\nabla\t_j(\tilde{x})-K^{-1}\nabla\phi_j(\tilde{x})\right) \\
+ \frac{\pi}{2}\left(K+K^{-1}\right)\d(0)\left(\d_{i+1,j}+\d_{ij}\right)\end{array}\right]\psi^\dg_{i\a}(\tilde{x})\psi_{i+1,\b}(x)
\end{equation}
\end{widetext}

\noindent There are four terms of this type in $[h'_i,h_j]$, and another four in $[h'_j,h_i]$.  Adding them all and summing over $i$, then using the trick of shifting the chain index $j$ in half the terms before taking the limit $k\rightarrow 0$ as in equation \eqref{eq:index_k_to_0_trick}, gives the thermal current operator, equation \eqref{eq:LL_J_E}.

\subsection{Thermal current-current correlator}
The thermal expectation value $\langle J_E(\tau)J_E\rangle$ looks like $P\int dx dx' \sum_j \langle \cdots \rangle$ where $P$ is some (dimensionful) prefactor, four integrals over real-space coordinates have been reduced to two by assuming $t(x-x')\propto\d(x-x')$ (see Appendix \ref{appendix:correspondence}), and the expectation value is a sum of terms of the form 
\begin{equation}
\langle [\nabla_i]_x\psi_j^\dg(x)\psi_j(x)[\nabla_{i'}]_{x'}\psi_{j'}^\dg(x')\psi_{j'}(x')\rangle\label{eq:JEJE_term_form}
\end{equation} 

\noindent where the $[\nabla]$ operators are defined in equation \eqref{eq:def_nabla}.  The indices satisfy $j'=j\pm 1$, with $i$ and $i'$ related to $j$ and $j'$ in one of four possible ways; these four cases are: (1) $i=i'=j$, (2) $i=i'=j'$, (3) $i=j$ and $i'=j'$, and (4) $i=j'$ and $i'=j$.  As in the generalized noninteracting model, the fact that we work only to lowest order in the interchain hopping allows us to drop the hopping terms from the Hamiltonian appearing in the density matrix used in the calculation of the expectation values, $e^{-\b H}\rightarrow e^{-\b H_0}$, and likewise for the time evolution, so that the expectation values for each term of the type in equation \eqref{eq:JEJE_term_form} splits up into a product of expectation values on two individual chains.  Cases (1) through (4) lead to eight different types of two-point functions on the individual chains, as follows:
\begin{subequations}
\begin{align}
(1) & \rightarrow \langle [\nabla]\psi^\dg[\nabla]\psi\rangle \langle \psi\psi^\dg\rangle\\
(2) & \rightarrow \langle \psi^\dg\psi\rangle \langle [\nabla]\psi[\nabla]\psi^\dg\rangle\\
(3) & \rightarrow \langle [\nabla]\psi^\dg\psi\rangle \langle \psi[\nabla]\psi^\dg\rangle\\
(4) & \rightarrow \langle \psi^\dg[\nabla]\psi\rangle \langle [\nabla]\psi\psi^\dg\rangle
\end{align}
\end{subequations}

\noindent Both $\langle \psi_\a(x,\tau)\psi)^\dg_\a(0,0)\rangle$ and $\langle \psi^\dg_\a(x,\tau)\psi_\a(0,0)\rangle$ can be written simply in terms of the single-chain Green's function, being $-G_\a(x,\tau)$ and $G_\a(-x,-\tau)$ respectively; these are the only two that appear in the calculation of $\langle J(\tau)J\rangle$ and therefore in the calculation of the electrical conductivity.

The other six types of two-point functions we compute by writing them in terms of derivatives of the Green's function.  The first step is to separate the $[\nabla]$ operator into two pieces, proportional to $\a\phi-\theta$ and $-\a\phi-\theta$, 
\begin{equation}
[\nabla_j]^\a_y = -\a\nabla_y \left[\g(\a\phi_j-\t_j) + \tilde{\g}(-\a\phi_j-\t_j)\right]\label{eq:box_nabla_split}
\end{equation}

\noindent where $\g=(K+K^{-1})/2$ as usual and $\tilde{\g}=(K-K^{-1})/2$.  This operator only appears in expectation values with $\psi_\a$ and $\psi^\dg_\a$, which according to equations \eqref{eq:LL_psi_from_pt} contain $\a\phi-\theta$ but not $-\a\phi-\theta$.  Then when $[\nabla]$ is split up inside an expectation value and the expectation values of the two terms are calculated separately, all of the $-\a\phi-\theta$ terms vanish.  (See further discussion of this point in the Supplemental Material\cite{SuppMat}.) 

A factor of $\a\nabla\phi-\nabla\theta$ is pulled down by every derivative of $\psi_\a$ or $\psi^\dg_\a$, so that for instance
\begin{widetext}
\begin{equation}
\langle [\nabla]^\a_{x,\tau} \psi_\a(x,\tau)\psi^\dg_\a(x')\rangle = \a\g\langle i e^{i\a k_F x}\nabla_x(e^{-i\a k_F x}\psi_\a(x,\tau))\psi^\dg_\a(x')\rangle = -i\a\g e^{i\a k_F x,\tau}\nabla_x \left(e^{-i\a k_F x}G_\a(x-x',\tau)\right).
\end{equation}
The remaining five two-point functions are calculated in a similar manner.  For cases (1) through (4) we find 
\begin{subequations}
\begin{align}
\langle [\nabla]\psi^\dg[\nabla]\psi\rangle \langle \psi\psi^\dg\rangle & = \g^2 \left[(\a k_F)^2 G_\a(x-x',\tau) + 2i\a k_F \partial_x G_\a(x-x',\tau) - \partial^2_x G_\a(x-x',\tau)\right]\tilde{G}_\a(x-x',\tau)\\
\langle \psi^\dg\psi\rangle \langle [\nabla]\psi[\nabla]\psi^\dg\rangle & = \g^2 \left[(\a k_F)^2 \tilde{G}_\a(x-x',\tau) - 2i\a k_F \partial_x \tilde{G}_\a(x-x',\tau) - \partial^2_x \tilde{G}_\a(x-x',\tau)\right]G_\a(x-x',\tau)\\
\langle [\nabla]\psi^\dg\psi\rangle \langle \psi[\nabla]\psi^\dg\rangle & = \g^2 \left[(k_F-i\a\partial_x)\tilde{G}_\a(x-x',\tau)\right]\times\left[(k_F+i\a\partial_x)G_\a(x-x',\tau)\right]\label{eq:npppnp}\\
\langle \psi^\dg[\nabla]\psi\rangle \langle [\nabla]\psi\psi^\dg\rangle & = \g^2 \left[(k_F-i\a\partial_x)\tilde{G}_\a(x-x',\tau)\right]\times\left[(k_F+i\a\partial_x)G_\a(x-x',\tau)\right]\label{eq:pnpnpp}
%
%\langle[\nabla_j]^\a_{x,\tau}\psi^\dg_{j\a}(x,\tau)\psi_{j\a}(x')\rangle\langle\psi_{j-1,\a}(x,\tau)[\nabla_{j-1}]^\a_{x'}\psi^\dg_{j-1,\a}(x')\rangle & = \g^2 \left[(k_F-i\a\partial_x)\tilde{G}_\a(x-x',\tau)\right]\times\left[(k_F+i\a\partial_x)G_\a(x-x',\tau)\right]\\
%
%\langle[\nabla_{j-1}]^\a_{x,\tau}\psi_{j-1,\a}(x,\tau)\psi^\dg_{j-1,\a}(x')\rangle\langle\psi^\dg_{j\a}(x,\tau)[\nabla_j]^\a_{x'}\psi_{j\a}(x')\rangle & = \g^2 \left[(k_F+i\a\partial_x)G_\a(x-x',\tau)\right]\times\left[(k_F-i\a\partial_x)\tilde{G}_\a(x-x',\tau)\right]\\
%
%\langle[\nabla_{j-1}]^\a_{x,\tau}\psi_{j-1,\a}(x,\tau)[\nabla_{j-1}]^\a_{x'}\psi^\dg_{j-1,\a}(x')\rangle\langle\psi^\dg_{j\a}(x,\tau)\psi_{j\a}(x')\rangle & = \g^2 \left[(\a k_F)^2 G_\a(x-x',\tau) + 2i\a k_F \partial_x G_\a(x-x',\tau) - \partial^2_x G_\a(x-x',\tau)\right]\tilde{G}_\a(x-x',\tau)\\
%
%\langle[\nabla_j]^\a_{x,\tau}\psi^\dg_{j\a}(x,\tau)[\nabla_j]^\a_{x'}\psi_{j\a}(x')\rangle\langle\psi_{j-1,\a}(x,\tau)\psi^\dg_{j-1,\a}(x')\rangle & = \g^2 \left[(\a k_F)^2 \tilde{G}_\a(x-x',\tau) - 2i\a k_F \partial_x \tilde{G}_\a(x-x',\tau) - \partial^2_x \tilde{G}_\a(x-x',\tau)\right]G_\a(x-x',\tau)
\end{align}
\end{subequations}
\end{widetext}

\noindent for $\tilde{G}(x,\tau)=-G(-x,-\tau)$.  We have omitted indices and coordinates on the left-hand side for clarity.  The last two terms are clearly the same, but the first two appear to be different.  In fact, all of these expressions are inside an integral over $x$ and $x'$, so we apply integration by parts to move derivatives in the first two terms; the result is that all four terms are equal.  These expressions, for instance in equation \eqref{eq:npppnp}, are now quite reminiscent of equation \eqref{eq:LL_JEJECorrelator_unspecified_Greens} for $\langle J_E(\tau)J_E\rangle$ in the main paper.

To finish the calculation, we change variables in the integration from $x$ and $x'$ to $x-x'$ and $(x+x')/2$.  The integrand does not depend on the center of mass coordinate and thus the integral over $(x+x')/2$ just provides a factor of the length of the 1D chain.  The result is equation \eqref{eq:LL_JEJECorrelator_unspecified_Greens}.

\begin{widetext}

\subsection{Correlator results in terms of \texorpdfstring{$F_1$}{F1}}

By substituting the Luttinger liquid Green's function, equation \eqref{eq:LL_Greens}, into the current-current correlators, equations \eqref{eq:JJCorrelators_unspecified_Greens}, and integrating over the position $x$ from $-\infty$ to $\infty$, we find expressions for the correlators that are functions only of the imaginary time $\tau$.  In practice we write the results in terms of the dimensionless parameter $\tau'=\tau\pi/\b$ because that makes it easy to separate the dimensionful parts of the transport coefficients as given in equations \eqref{eq:LL_transport_dimensions} from the purely numerical parts that we need only for finding the Lorenz number.

The expression for $\langle J(\tau')J\rangle$ is given in equation \eqref{eq:LL_JeJeCorrelatorFull} in the main paper. The corresponding expressions for the remaining two correlators are
%\begin{widetext}
\begin{subequations}
\begin{align}
\langle J_E(\tau') J_E\rangle & = N_cL\g^2\left(\frac{a_c v t}{2\pi}\right)^2\frac{1}{2a^3\pi^2}\left(\frac{2\pi a}{v\b}\right)^{2\g+1}\label{eq:LL_JEJECorrelatorFull}
\times
\left(
\begin{array}{c}
-4 (2 + \g^2 - 2\cos(4\tau')) f(\g,\tau',3,3)\\\\
+ \cos(2\tau') (2 + \g^2 - 2\cos(4\tau'))(f(\g,\tau',2,3) + f(\g,\tau',4,3))\\\\
+ 2(1 + \g^2 - \cos(4\tau'))(f(\g,\tau',1,3) + f(\g,\tau',5,3))\\\\
 - \g^2 \cos(2\tau') (f(\g,\tau',0,3) + f(\g,\tau',6,3))
\end{array}\right)\\
\langle J_E(\tau') J_E\rangle & = 2v\g N_c L\left(\frac{a_c t}{2\pi}\right)^2 \frac{1}{a^2\pi^2}\left(\frac{2\pi a}{v\b}\right)^{2\g}\sin(2\tau')\label{eq:LL_JEJeCorrelatorFull}
\times
\left(
\begin{array}{c}
-2 (1 + \frac{\g}{2})f(\g,\tau',2,2)\\\\
+ \cos(2\tau')(f(\g,\tau',1,2) + f(\g,\tau',3,2))\\\\
+ \frac{\g}{2}(f(\g,\tau',0,2) + f(\g,\tau',4,2))
\end{array}\right)
\end{align}
\end{subequations}
\end{widetext}

\noindent The function $f(\g,\tau,n,m)$ can be written in terms of the Appell hypergeometric function $F_1$ as in equation \eqref{eq:f_from_F1} in the main paper, and it also has a nice integral representation, 
\begin{equation}
f(\g,\tau,n,m) = \int_0^1 t^{\g+n-1}(1-2t\cos(2\tau)+t^2)^{-(\g+m)}\,dt,
\end{equation} 
which is derived in the Supplemental Material\cite{SuppMat} from a representation of this type for $F_1$.

\subsection{Numerical Fourier transform and analytic continuation} 

Computing the $L^{(il)}$ coefficients involves evaluating the expression 
\begin{equation}
\lim_{\omega\rightarrow 0}\lim_{\d\rightarrow 0}\frac{1}{\omega}\left[\int_0^\b e^{i\omega_n\tau}\langle j_l(\tau) j_i\rangle \, d\tau\right]_{i\omega_n\rightarrow \omega+i\d}.
\end{equation}

\noindent The first step is to write anything that cannot be computed analytically in terms of dimensionless quantities, which we do by the transformation $\tau\rightarrow\tau'$.  This results in
\begin{equation}
\lim_{n\rightarrow 0}\lim_{\d'\rightarrow 0}\frac{\b^2}{2\pi^2 n}\left[\int_0^\pi e^{2in\tau'}\langle j_l(\tau') j_i\rangle \, d\tau'\right]_{in\rightarrow n+i\d'}.\label{eq:FT_and_lim_nondim}
\end{equation}

\noindent In principle we would now find a unique analytic function $f(n)$ such that $\int e^{2in\tau'}\langle j(\tau') j\rangle d\tau' = f(n)$ for every $n=0,1,2,\cdots$, but there is no general formula for the Fourier transforms and the integrals must therefore be computed individually for each value of $n$.  This provides a limited set of points $(n,f(n))$ to use in fitting an analytic function.  

Two standard approaches to this function-fitting problem are the maximum entropy method\cite{Jarrell1996,Gunnarsson2010} and the Pad\'{e} approximation\cite{Vidberg1977,Pade_ref}.  The maximum entropy method is more robust to numerical errors, but it does depend quite strongly on an initial assumption of the form of the function.  In our case, we do not a priori have any strong assumptions about what the function $f(n)$ should look like, and our data comes from numerical integrals for which we can bound the error by requiring a fixed level of precision, with no statistical errors like those that appear in quantum Monte Carlo calculations.  We therefore use the Pad\'{e} approximation and fit a rational function to the calculated Fourier transforms at Matsubara frequencies.

If we evaluate the Fourier transform at $2N$ points, we can find an exact fit for a rational function with $2N$ parameters, namely
\begin{equation}
f(x)=\frac{\sum_{n=1}^{N} a_n x^n}{\sum_{n=0}^{N+1} b_n x^n}.
\end{equation}
This has only $2N$ parameters because $f(0)$ is just the integral of the current-current correlation function so that $b_0\neq 0$, and therefore we can assume without loss of generality that $b_0=1$.  Our method for finding $f$ from the $2N$ points is discussed further in the Supplemental Material\cite{SuppMat} and is very similar to the method described in reference \onlinecite{Pade_ref}.

A major benefit of writing $f(x)$ as a rational function is that the analytic continuation can be accomplished simply by the replacement $n\rightarrow \d'-in$.  We make this substitution, divide by $n$ (from equation \ref{eq:FT_and_lim_nondim}), and take the imaginary part to get only the real part of $L^{(il)}$; letting both $n$ and $\d'$ go to 0, we find in the case that the correlation function $\langle j_l(\tau')j_i\rangle$ is even about $\tau'=\pi/2$ the very simple expression
\begin{equation}
\lim_{n\rightarrow 0}\lim_{\d'\rightarrow 0}\,\left(\text{Im}\left[\frac{f(\d'-in)}{n}\right]\right)=a_0 b_1-a_1
\end{equation}
which is just minus the derivative of $f(x)$ evaluated at $x=0$.  (Note that under some assumptions about $f$, this follows from the Cauchy-Riemann equations.)  If the correlation function is odd about $\tau'=\pi/2$, then we get 0.

It turns out that the function $f(\g,\tau,n,m)$ is even about $\tau'=\pi/2$, which implies that both $\langle J(\tau)J\rangle$ (equation \ref{eq:LL_JeJeCorrelatorFull}) and $\langle J_E(\tau)J_E\rangle$ (equation \ref{eq:LL_JEJECorrelatorFull}) are even, while $\langle J_E(\tau)J\rangle$ (equation \ref{eq:LL_JEJeCorrelatorFull}) is odd.  This is the mathematical explanation for why the thermopower vanishes in our calculation for the Luttinger liquid model, although of course this result was expected due to particle-hole symmetry.

There are two complications that must be addressed.  First, the form of the function $f(x)$ and hence the calculated value for the numerical part of $L^{(il)}$ depends on the number of points used to fit the function.  With a small number of points, the function is highly underdetermined and thus the derivative at the origin is inaccurate.  Conversely, finding the parameters in $f$ involves inverting a matrix that quickly becomes ill-conditioned as $N$ grows, which for a given precision of the numerical integrals sets an upper bound on how many data points we can use.  In practice, we compute the transport coefficients for every value of $N$ from 1 through $N_\text{max}$, confirm that the resulting numerical series converges, and use the limit of the sequence for the value of the transport coefficient.  We use $N_\text{max}=40$ because that value empirically gives good convergence for all transport coefficients that we calculate.

The second complication is that the functions $f(\g,\tau,n,m)$ are divergent at $\tau=0$ and $\pi$.  We regulate the divergence by introducing a cutoff $\epsilon$ at both bounds of the integral in equation \eqref{eq:FT_and_lim_nondim}, integrating from $\epsilon$ to $\pi-\epsilon$ instead of $0$ to $\pi$.  We compute the transport coefficients for values of $\epsilon$ that vary over an order of magnitude (from $0.1$ to $0.01$) and confirm that the results for the transport coefficients converge as $\epsilon\rightarrow 0$.  The numerical error grows as $\epsilon\rightarrow 0$, so all the numerical results for the Luttinger liquid model shown in figures \ref{fig:combined_Lorenz_number} and \ref{fig:LL_L_exponent} are for $\epsilon=10^{-1.5}$, for which the results are converged and the error is guaranteed to be small.  See the Supplemental Material\cite{SuppMat} for details.

%=======================================================================
%=======================================================================

\section{Correspondence between the two models\label{appendix:correspondence}}

In the main text of the paper we have compared the results of our two models, implicitly assuming that the results they give should match at least in the noninteracting limit.  In this appendix we confirm that the two models match in that limit, first by showing that the hopping terms in the two models are equivalent and second by explicitly rewriting the Fourier-space expression for $\langle J(\tau)J\rangle$ from the generalized noninteracting model in a real-space representation and showing that the result matches the noninteracting limit of equation \eqref{eq:LL_JeJeCorrelator_unspecified_Greens} from the Luttinger liquid model.

\subsection{Correspondence of hopping terms}

The correspondence between the Fourier-space operators $c_k$ that appear in the Hamiltonian in equation \eqref{eq:nonint_H} and the real-space operators $\psi_\a(x)$ that appear in the Hamiltonian in equation \eqref{eq:LL_model_H} is given by a Fourier transform,
\begin{subequations}
\begin{align}
c_{k\a} & = \frac{1}{\sqrt L}\int e^{-ikx}e^{-i\a k_F x}\psi_\a(x) dx\label{eq:psi_to_c_split}\\
\psi_\a(x) &= \frac{e^{i\a k_F x}}{\sqrt L}\sum_k e^{ikx}c_{k\a}\label{eq:c_to_psi_split}
\end{align}\label{eq:Fourier_conversion_general}
\end{subequations}\

\noindent where the chiral Fourier-space operator $c^\dg_{k\a}$ creates a fermion that has wave-vector $k$ relative to the Fermi point $\a k_F$.  We can then find the correspondence between the hopping strength $t_{kk'}$ from equation \eqref{eq:nonint_H} and $t_{\a\b}(x-x')$ from equation \eqref{eq:LL_model_H} by substituting equation \eqref{eq:c_to_psi_split} into the hopping term of the Luttinger liquid Hamiltonian and matching the result to the corresponding term in the noninteracting Hamiltonian.  To simplify the calculation, we rewrite the hopping part of the Luttinger liquid with only two terms, as 
\begin{equation}
\sum_{\a\b}\int dx\,dx'\left[t_{\a\b}(x-x')\psi_{j\a}^\dg(x)\psi_{j+1,\b}(x') + \text{h.c.}\right].
\end{equation}

\noindent In fact, it is sufficient to match just the first term of this to the first term in the hopping part of the noninteracting Hamiltonian, since their Hermitian conjugates will automatically match as well.

Making the substitution with equation \eqref{eq:c_to_psi_split}, we have:
\begin{widetext}
\begin{subequations}
\begin{align}
\sum_{\a\b}\!\int\! dx\,dx'\, t_{\a\b}(x-x')\psi_{j\a}^\dg(x)\psi_{j+1,\b}(x') & = \sum_{\a\b}\!\int\! dx\,dx'\left[t_{\a\b}(x-x')\!\!\left[\!\frac{e^{-i\a k_F x}}{\sqrt L}\sum_k e^{-ikx}c^\dg_{jk\a}\!\right]\!\!\left[\!\frac{e^{i\b k_F x'}}{\sqrt L}\sum_k e^{ikx'}c_{j+1,k\b}\!\right]\right]\\
& = \frac{1}{L}\sum_{kk'}\sum_{\a\b}\int \!dx\,dx'\left[t_{\a\b}(x-x')e^{-i k_F (\a x-\b x')}\left[e^{-ikx}e^{ik'x'}\right]c^\dg_{jk\a}c_{j+1,k'\b}\right]
\end{align}
\end{subequations}
\end{widetext}

\noindent We can compare this with the equivalent term for the generalized noninteracting model, which looks like
\begin{equation}
\sum_{kk'} t_{kk'}c^\dg_{j,k} c_{j+1,k'} = \sum_{kk'}\sum_{\a\b} t_{kk'}\d_{\a\b}c^\dg_{jk\a} c_{j+1,k'\b}.
\end{equation}

\noindent For the two to be equal, we must have $t_{\a\b}(x-x')=\d_{\a\b}t(x-x')$ and 
\begin{equation}
t_{kk'} = \frac{1}{L}\int dx\,dx'\left[t(x-x')e^{-i\a k_F (x-x')}e^{-ikx}e^{ik'x'}\right].\label{eq:hopping_fourier_x_to_k}
\end{equation}

\noindent The inverse relation is 
\begin{equation}
t(x-x')e^{-i\a k_F (x-x')} = \frac{L}{(2\pi)^2}\int dk\,dk'\, t_{kk'} e^{ikx}e^{-ik'x'}.
\label{eq:hopping_fourier_k_to_x}
\end{equation}

\noindent From these relations, we can verify the consistency of the hopping strengths that we used in our calculations, namely $t(k,k')=(t/L)\d(k-k')$ from equation \eqref{eq:tkk} and $t(x-x')\propto\d(x-x')$.  Starting from $t(k,k')$ and using equation \eqref{eq:hopping_fourier_k_to_x}, we find 
\begin{equation}
t(x-x')=\frac{t}{2\pi}\d(x-x').
\end{equation}

\noindent Note that the factor of $L^{-1}$ in $t(k,k')$ is necessary to cancel the factor of $L$ in equation \eqref{eq:hopping_fourier_k_to_x}, so that the hopping strength $t(x-x')$ between localized sites does not depend on the chain length; such a dependence would be unphysical.

The factor of $L^{-1}$ in front of the delta function in $t(k,k')$ appears because the width of the Gaussian describing $t_{kk'}$ is proportional to $L^{-1}$.  We assume the specific form of the hopping $t_{kk'}$ given in equation \eqref{eq:tkk} specifically to achieve the cancellation of factors of the length of the system in $t(x,x')$.  This ensures that both $t(x-x')$ and $t_{kk'}$ are physically valid, while also being compatible with each other according to equations \eqref{eq:hopping_fourier_x_to_k} and \eqref{eq:hopping_fourier_k_to_x}.

\begin{widetext}
%\onecolumngrid

\subsection{Real space representation of current-current correlator in generalized noninteracting model}

In the noninteracting limit, $\g\rightarrow 1$, the results of our two models should precisely agree.  We confirm that explicitly by writing $\langle J(\tau)J\rangle$ as calculated in the generalized noninteracting model in a real-space representation.  We begin from equation \eqref{eq:nonint_JeJe_intermediate}, first converting back into an integral over $k$ to get
%\begin{widetext}
\begin{equation}
\langle J(\tau)J\rangle = 2N_c a_c^2 \left(\frac{L}{2\pi}\right)^2\sum_\a\int_{kk'}|t(k,k')|^2 \left[\frac{e^{\tau(E_\a(k)-E_\a(k'))}}{\left(1+e^{\b E_\a(k)}\right)\left(1+e^{-\b E_\a(k')}\right)}\right]dk\,dk'
\end{equation}

\noindent where on each branch ($\a=R,L$), $k$ is measured from the Fermi point $\a k_F$.  Putting in the linear dispersion $E_\a(k) = \a v k$ and substituting equation \eqref{eq:hopping_fourier_x_to_k} for $t(k,k')$, this becomes
\begin{align}
\langle J(\tau)J\rangle & = \frac{2N_c a_c^2}{(2\pi)^2} \sum_\a\int \begin{array}{c}dx_1\,dx_2\\dx_3\,dx_4\end{array}\left[t(x_1-x_2)e^{-i\a k_F (x_1-x_2)}\right]\left[t(x_3-x_4)^\ast e^{i\a k_F (x_3-x_4)}\right]\nonumber\\
& \,\,\,\,\,\,\,\,\,\,\,\,\,\,\,\,\,\,\,\,\,\,\,\,\,\,\,\,\,\,\,\,\,\,\,\,\,\,\,\,\,\,\,\,\,\,\,\,\,\,\,\,\,\,\,\,\,\,\,\,\,\,\,\,\,\,\,\,\,\,\,\,\,\,\,\,\,\,\,\, \times \left[\int dk\,\frac{e^{\a\tau'\b vk/\pi}e^{-ik(x_1-x_3)}}{1+e^{\a\b vk}}\right]\left[\int dk'\,\frac{e^{-\a\tau'\b v k'/\pi}e^{ik'(x_2-x_4)}}{1+e^{-\a\b vk'}}\right]
\end{align}

\noindent Substituting $t(x-x')=(t/2\pi)\d(x-x')$ and computing the integrals over $k$ and $k'$ gives
\begin{align}
\langle J(\tau)J\rangle & = \frac{2N_c a_c^2 t^2}{(2\pi)^4} \sum_\a\int dx\,dx\,\left[-\frac{i\pi}{v\b}\text{csch}\left(\frac{\pi}{v\b}(x'-x-i\tau)\right)\right]\left[\frac{i\pi}{v\b}\text{csch}\left(\frac{\pi}{v\b}(x-x'+i\tau)\right)\right]\\
& = -\frac{4N_cLa_c^2t^2}{(2\pi)^4} \left(\frac{\pi}{v\b}\right)^2\int dx\,\left[\text{csch}\left(\frac{\pi}{v\b}(x+i\tau)\right)\right]^2\label{eq:nonint_JJ_correlator_realspace}
\end{align}

\noindent This result can be compared with the noninteracting ($\g=1$) limit of $\langle J(\tau)J\rangle$ in the Luttinger liquid model, as given by equation \eqref{eq:LL_JeJeCorrelator_unspecified_Greens}.  The noninteracting Green's function is found by substituting $\g=1$ into equation \eqref{eq:LL_Greens} to get
\begin{equation}
\lim_{\g\rightarrow 1}G_\a(x,\tau) = -\frac{e^{i\a k_F x}}{2\pi}\left[\frac{i\a}{\frac{v\b}{\pi}\sinh\left(\frac{x+i\a v\tau}{v\b/\pi}\right)}\right],\label{eq:LL_Greens_nonint}
\end{equation}
and substituting this into equation \eqref{eq:LL_JeJeCorrelator_unspecified_Greens} gives
\begin{equation}
\langle J(\tau)J\rangle = -\frac{2N_cL a_c^2 t^2}{(2\pi)^4} \left(\frac{\pi}{v\b}\right)^2\sum_\a\int dx\,\left[\text{csch}\left(\frac{\pi}{v\b}(x+i\a\tau)\right)\right]^2
\end{equation}
\end{widetext}

\noindent The integral does not actually depend on $\a$ since all terms containing $\a$ are odd in $x$ and integrate to 0.  We can therefore let $\a\rightarrow 1$ in the integrand, in which case the sum over $\a$ becomes just a factor of 2 and the result precisely matches the real-space representation of the correlator from the generalized noninteracting model, equation \eqref{eq:nonint_JJ_correlator_realspace}.


\begin{thebibliography}{57}%
\makeatletter
\providecommand \@ifxundefined [1]{%
 \@ifx{#1\undefined}
}%
\providecommand \@ifnum [1]{%
 \ifnum #1\expandafter \@firstoftwo
 \else \expandafter \@secondoftwo
 \fi
}%
\providecommand \@ifx [1]{%
 \ifx #1\expandafter \@firstoftwo
 \else \expandafter \@secondoftwo
 \fi
}%
\providecommand \natexlab [1]{#1}%
\providecommand \enquote  [1]{``#1''}%
\providecommand \bibnamefont  [1]{#1}%
\providecommand \bibfnamefont [1]{#1}%
\providecommand \citenamefont [1]{#1}%
\providecommand \href@noop [0]{\@secondoftwo}%
\providecommand \href [0]{\begingroup \@sanitize@url \@href}%
\providecommand \@href[1]{\@@startlink{#1}\@@href}%
\providecommand \@@href[1]{\endgroup#1\@@endlink}%
\providecommand \@sanitize@url [0]{\catcode `\\12\catcode `\$12\catcode
  `\&12\catcode `\#12\catcode `\^12\catcode `\_12\catcode `\%12\relax}%
\providecommand \@@startlink[1]{}%
\providecommand \@@endlink[0]{}%
\providecommand \url  [0]{\begingroup\@sanitize@url \@url }%
\providecommand \@url [1]{\endgroup\@href {#1}{\urlprefix }}%
\providecommand \urlprefix  [0]{URL }%
\providecommand \Eprint [0]{\href }%
\providecommand \doibase [0]{http://dx.doi.org/}%
\providecommand \selectlanguage [0]{\@gobble}%
\providecommand \bibinfo  [0]{\@secondoftwo}%
\providecommand \bibfield  [0]{\@secondoftwo}%
\providecommand \translation [1]{[#1]}%
\providecommand \BibitemOpen [0]{}%
\providecommand \bibitemStop [0]{}%
\providecommand \bibitemNoStop [0]{.\EOS\space}%
\providecommand \EOS [0]{\spacefactor3000\relax}%
\providecommand \BibitemShut  [1]{\csname bibitem#1\endcsname}%
\let\auto@bib@innerbib\@empty
%</preamble>
\bibitem [{\citenamefont {Yuen}\ \emph {et~al.}(2009)\citenamefont {Yuen},
  \citenamefont {Menon}, \citenamefont {Coates}, \citenamefont {Namdas},
  \citenamefont {Cho}, \citenamefont {Hannahs}, \citenamefont {Moses},\ and\
  \citenamefont {Heeger}}]{Yuen2009}%
  \BibitemOpen
  \bibfield  {author} {\bibinfo {author} {\bibfnamefont {J.~D.}\ \bibnamefont
  {Yuen}}, \bibinfo {author} {\bibfnamefont {R.}~\bibnamefont {Menon}},
  \bibinfo {author} {\bibfnamefont {N.~E.}\ \bibnamefont {Coates}}, \bibinfo
  {author} {\bibfnamefont {E.~B.}\ \bibnamefont {Namdas}}, \bibinfo {author}
  {\bibfnamefont {S.}~\bibnamefont {Cho}}, \bibinfo {author} {\bibfnamefont
  {S.~T.}\ \bibnamefont {Hannahs}}, \bibinfo {author} {\bibfnamefont
  {D.}~\bibnamefont {Moses}}, \ and\ \bibinfo {author} {\bibfnamefont {A.~J.}\
  \bibnamefont {Heeger}},\ }\href {\doibase 10.1038/nmat2470} {\bibfield
  {journal} {\bibinfo  {journal} {Nat Mater}\ }\textbf {\bibinfo {volume}
  {8}},\ \bibinfo {pages} {572} (\bibinfo {year} {2009})}\BibitemShut {NoStop}%
\bibitem [{\citenamefont {Kim}\ \emph {et~al.}(2013)\citenamefont {Kim},
  \citenamefont {Shao}, \citenamefont {Zhang},\ and\ \citenamefont
  {Pipe}}]{Pipe2013}%
  \BibitemOpen
  \bibfield  {author} {\bibinfo {author} {\bibfnamefont {G.-H.}\ \bibnamefont
  {Kim}}, \bibinfo {author} {\bibfnamefont {L.}~\bibnamefont {Shao}}, \bibinfo
  {author} {\bibfnamefont {K.}~\bibnamefont {Zhang}}, \ and\ \bibinfo {author}
  {\bibfnamefont {K.~P.}\ \bibnamefont {Pipe}},\ }\href {\doibase
  10.1038/nmat3635} {\bibfield  {journal} {\bibinfo  {journal} {Nat Mater}\
  }\textbf {\bibinfo {volume} {12}},\ \bibinfo {pages} {719} (\bibinfo {year}
  {2013})}\BibitemShut {NoStop}%
\bibitem [{\citenamefont {Bubnova}\ \emph {et~al.}(2014)\citenamefont
  {Bubnova}, \citenamefont {Khan}, \citenamefont {Wang}, \citenamefont {Braun},
  \citenamefont {Evans}, \citenamefont {Fabretto}, \citenamefont
  {Hojati-Talemi}, \citenamefont {Dagnelund}, \citenamefont {Arlin},
  \citenamefont {Geerts}, \citenamefont {Desbief}, \citenamefont {Breiby},
  \citenamefont {Andreasen}, \citenamefont {Lazzaroni}, \citenamefont {Chen},
  \citenamefont {Zozoulenko}, \citenamefont {Fahlman}, \citenamefont {Murphy},
  \citenamefont {Berggren},\ and\ \citenamefont {Crispin}}]{Bubnova2013}%
  \BibitemOpen
  \bibfield  {author} {\bibinfo {author} {\bibfnamefont {O.}~\bibnamefont
  {Bubnova}}, \bibinfo {author} {\bibfnamefont {Z.~U.}\ \bibnamefont {Khan}},
  \bibinfo {author} {\bibfnamefont {H.}~\bibnamefont {Wang}}, \bibinfo {author}
  {\bibfnamefont {S.}~\bibnamefont {Braun}}, \bibinfo {author} {\bibfnamefont
  {D.~R.}\ \bibnamefont {Evans}}, \bibinfo {author} {\bibfnamefont
  {M.}~\bibnamefont {Fabretto}}, \bibinfo {author} {\bibfnamefont
  {P.}~\bibnamefont {Hojati-Talemi}}, \bibinfo {author} {\bibfnamefont
  {D.}~\bibnamefont {Dagnelund}}, \bibinfo {author} {\bibfnamefont {J.-B.}\
  \bibnamefont {Arlin}}, \bibinfo {author} {\bibfnamefont {Y.~H.}\ \bibnamefont
  {Geerts}}, \bibinfo {author} {\bibfnamefont {S.}~\bibnamefont {Desbief}},
  \bibinfo {author} {\bibfnamefont {D.~W.}\ \bibnamefont {Breiby}}, \bibinfo
  {author} {\bibfnamefont {J.~W.}\ \bibnamefont {Andreasen}}, \bibinfo {author}
  {\bibfnamefont {R.}~\bibnamefont {Lazzaroni}}, \bibinfo {author}
  {\bibfnamefont {W.~M.}\ \bibnamefont {Chen}}, \bibinfo {author}
  {\bibfnamefont {I.}~\bibnamefont {Zozoulenko}}, \bibinfo {author}
  {\bibfnamefont {M.}~\bibnamefont {Fahlman}}, \bibinfo {author} {\bibfnamefont
  {P.~J.}\ \bibnamefont {Murphy}}, \bibinfo {author} {\bibfnamefont
  {M.}~\bibnamefont {Berggren}}, \ and\ \bibinfo {author} {\bibfnamefont
  {X.}~\bibnamefont {Crispin}},\ }\href {\doibase 10.1038/nmat3824} {\bibfield
  {journal} {\bibinfo  {journal} {Nat Mater}\ }\textbf {\bibinfo {volume}
  {13}},\ \bibinfo {pages} {190} (\bibinfo {year} {2014})}\BibitemShut
  {NoStop}%
\bibitem [{\citenamefont {Glaudell}\ \emph {et~al.}(2015)\citenamefont
  {Glaudell}, \citenamefont {Cochran}, \citenamefont {Patel},\ and\
  \citenamefont {Chabinyc}}]{Chabinyc2015}%
  \BibitemOpen
  \bibfield  {author} {\bibinfo {author} {\bibfnamefont {A.~M.}\ \bibnamefont
  {Glaudell}}, \bibinfo {author} {\bibfnamefont {J.~E.}\ \bibnamefont
  {Cochran}}, \bibinfo {author} {\bibfnamefont {S.~N.}\ \bibnamefont {Patel}},
  \ and\ \bibinfo {author} {\bibfnamefont {M.~L.}\ \bibnamefont {Chabinyc}},\
  }\href {\doibase 10.1002/aenm.201401072} {\bibfield  {journal} {\bibinfo
  {journal} {Advanced Energy Materials}\ }\textbf {\bibinfo {volume} {5}},\
  \bibinfo {pages} {1401072} (\bibinfo {year} {2015})}\BibitemShut {NoStop}%
\bibitem [{\citenamefont {Heeger}\ \emph {et~al.}(1988)\citenamefont {Heeger},
  \citenamefont {Kivelson}, \citenamefont {Schrieffer},\ and\ \citenamefont
  {Su}}]{Heeger1988}%
  \BibitemOpen
  \bibfield  {author} {\bibinfo {author} {\bibfnamefont {A.~J.}\ \bibnamefont
  {Heeger}}, \bibinfo {author} {\bibfnamefont {S.}~\bibnamefont {Kivelson}},
  \bibinfo {author} {\bibfnamefont {J.~R.}\ \bibnamefont {Schrieffer}}, \ and\
  \bibinfo {author} {\bibfnamefont {W.~P.}\ \bibnamefont {Su}},\ }\href
  {\doibase 10.1103/RevModPhys.60.781} {\bibfield  {journal} {\bibinfo
  {journal} {Rev. Mod. Phys.}\ }\textbf {\bibinfo {volume} {60}},\ \bibinfo
  {pages} {781} (\bibinfo {year} {1988})}\BibitemShut {NoStop}%
\bibitem [{\citenamefont {Sai}\ \emph {et~al.}(2007)\citenamefont {Sai},
  \citenamefont {Li}, \citenamefont {Martin}, \citenamefont {Basov},\ and\
  \citenamefont {Di~Ventra}}]{Sai2007}%
  \BibitemOpen
  \bibfield  {author} {\bibinfo {author} {\bibfnamefont {N.}~\bibnamefont
  {Sai}}, \bibinfo {author} {\bibfnamefont {Z.~Q.}\ \bibnamefont {Li}},
  \bibinfo {author} {\bibfnamefont {M.~C.}\ \bibnamefont {Martin}}, \bibinfo
  {author} {\bibfnamefont {D.~N.}\ \bibnamefont {Basov}}, \ and\ \bibinfo
  {author} {\bibfnamefont {M.}~\bibnamefont {Di~Ventra}},\ }\href {\doibase
  10.1103/PhysRevB.75.045307} {\bibfield  {journal} {\bibinfo  {journal} {Phys.
  Rev. B}\ }\textbf {\bibinfo {volume} {75}},\ \bibinfo {pages} {045307}
  (\bibinfo {year} {2007})}\BibitemShut {NoStop}%
\bibitem [{\citenamefont {Haldane}(1981)}]{Haldane1981Luttinger}%
  \BibitemOpen
  \bibfield  {author} {\bibinfo {author} {\bibfnamefont {F.~D.~M.}\
  \bibnamefont {Haldane}},\ }\href
  {http://stacks.iop.org/0022-3719/14/i=19/a=010} {\bibfield  {journal}
  {\bibinfo  {journal} {Journal of Physics C: Solid State Physics}\ }\textbf
  {\bibinfo {volume} {14}},\ \bibinfo {pages} {2585} (\bibinfo {year}
  {1981})}\BibitemShut {NoStop}%
\bibitem [{\citenamefont {Giamarchi}(2003)}]{GiamarchiBook}%
  \BibitemOpen
  \bibfield  {author} {\bibinfo {author} {\bibfnamefont {T.}~\bibnamefont
  {Giamarchi}},\ }\href@noop {} {\emph {\bibinfo {title} {Quantum Physics in
  One Dimension}}}\ (\bibinfo  {publisher} {Clarendon Press},\ \bibinfo
  {address} {Oxford},\ \bibinfo {year} {2003})\BibitemShut {NoStop}%
\bibitem [{\citenamefont {Segovia}\ \emph {et~al.}(1999)\citenamefont
  {Segovia}, \citenamefont {Purdie}, \citenamefont {Hengsberger},\ and\
  \citenamefont {Baer}}]{Segovia1999}%
  \BibitemOpen
  \bibfield  {author} {\bibinfo {author} {\bibfnamefont {P.}~\bibnamefont
  {Segovia}}, \bibinfo {author} {\bibfnamefont {D.}~\bibnamefont {Purdie}},
  \bibinfo {author} {\bibfnamefont {M.}~\bibnamefont {Hengsberger}}, \ and\
  \bibinfo {author} {\bibfnamefont {Y.}~\bibnamefont {Baer}},\ }\href {\doibase
  10.1038/990052} {\bibfield  {journal} {\bibinfo  {journal} {Nature}\ }\textbf
  {\bibinfo {volume} {402}},\ \bibinfo {pages} {504} (\bibinfo {year}
  {1999})}\BibitemShut {NoStop}%
\bibitem [{\citenamefont {Braunecker}\ \emph {et~al.}(2012)\citenamefont
  {Braunecker}, \citenamefont {Bena},\ and\ \citenamefont
  {Simon}}]{Braunecker}%
  \BibitemOpen
  \bibfield  {author} {\bibinfo {author} {\bibfnamefont {B.}~\bibnamefont
  {Braunecker}}, \bibinfo {author} {\bibfnamefont {C.}~\bibnamefont {Bena}}, \
  and\ \bibinfo {author} {\bibfnamefont {P.}~\bibnamefont {Simon}},\ }\href
  {\doibase 10.1103/PhysRevB.85.035136} {\bibfield  {journal} {\bibinfo
  {journal} {Phys. Rev. B}\ }\textbf {\bibinfo {volume} {85}},\ \bibinfo
  {pages} {035136} (\bibinfo {year} {2012})}\BibitemShut {NoStop}%
\bibitem [{\citenamefont {Blumenstein}\ \emph {et~al.}(2011)\citenamefont
  {Blumenstein}, \citenamefont {Schafer}, \citenamefont {Mietke}, \citenamefont
  {Meyer}, \citenamefont {Dollinger}, \citenamefont {Lochner}, \citenamefont
  {Cui}, \citenamefont {Patthey}, \citenamefont {Matzdorf},\ and\ \citenamefont
  {Claessen}}]{Blumenstein2011}%
  \BibitemOpen
  \bibfield  {author} {\bibinfo {author} {\bibfnamefont {C.}~\bibnamefont
  {Blumenstein}}, \bibinfo {author} {\bibfnamefont {J.}~\bibnamefont
  {Schafer}}, \bibinfo {author} {\bibfnamefont {S.}~\bibnamefont {Mietke}},
  \bibinfo {author} {\bibfnamefont {S.}~\bibnamefont {Meyer}}, \bibinfo
  {author} {\bibfnamefont {A.}~\bibnamefont {Dollinger}}, \bibinfo {author}
  {\bibfnamefont {M.}~\bibnamefont {Lochner}}, \bibinfo {author} {\bibfnamefont
  {X.~Y.}\ \bibnamefont {Cui}}, \bibinfo {author} {\bibfnamefont
  {L.}~\bibnamefont {Patthey}}, \bibinfo {author} {\bibfnamefont
  {R.}~\bibnamefont {Matzdorf}}, \ and\ \bibinfo {author} {\bibfnamefont
  {R.}~\bibnamefont {Claessen}},\ }\href {\doibase 10.1038/nphys2051}
  {\bibfield  {journal} {\bibinfo  {journal} {Nat Phys}\ }\textbf {\bibinfo
  {volume} {7}},\ \bibinfo {pages} {776} (\bibinfo {year} {2011})}\BibitemShut
  {NoStop}%
\bibitem [{\citenamefont {{Moser, J.}}\ \emph {et~al.}(1998)\citenamefont
  {{Moser, J.}}, \citenamefont {{Gabay, M.}}, \citenamefont {{Auban-Senzier,
  P.}}, \citenamefont {{Jérome, D.}}, \citenamefont {{Bechgaard, K.}},\ and\
  \citenamefont {{Fabre, J. M.}}}]{Moser1998}%
  \BibitemOpen
  \bibfield  {author} {\bibinfo {author} {\bibnamefont {{Moser, J.}}}, \bibinfo
  {author} {\bibnamefont {{Gabay, M.}}}, \bibinfo {author} {\bibnamefont
  {{Auban-Senzier, P.}}}, \bibinfo {author} {\bibnamefont {{Jérome, D.}}},
  \bibinfo {author} {\bibnamefont {{Bechgaard, K.}}}, \ and\ \bibinfo {author}
  {\bibnamefont {{Fabre, J. M.}}},\ }\href {\doibase 10.1007/s100510050150}
  {\bibfield  {journal} {\bibinfo  {journal} {Eur. Phys. J. B}\ }\textbf
  {\bibinfo {volume} {1}},\ \bibinfo {pages} {39} (\bibinfo {year}
  {1998})}\BibitemShut {NoStop}%
\bibitem [{\citenamefont {Wakeham}\ \emph {et~al.}(2011)\citenamefont
  {Wakeham}, \citenamefont {Bangura}, \citenamefont {Xu}, \citenamefont
  {Mercure}, \citenamefont {Greenblatt},\ and\ \citenamefont
  {Hussey}}]{Wakeham2011}%
  \BibitemOpen
  \bibfield  {author} {\bibinfo {author} {\bibfnamefont {N.}~\bibnamefont
  {Wakeham}}, \bibinfo {author} {\bibfnamefont {A.~F.}\ \bibnamefont
  {Bangura}}, \bibinfo {author} {\bibfnamefont {X.}~\bibnamefont {Xu}},
  \bibinfo {author} {\bibfnamefont {J.-F.}\ \bibnamefont {Mercure}}, \bibinfo
  {author} {\bibfnamefont {M.}~\bibnamefont {Greenblatt}}, \ and\ \bibinfo
  {author} {\bibfnamefont {N.~E.}\ \bibnamefont {Hussey}},\ }\href {\doibase
  10.1038/ncomms1406} {\bibfield  {journal} {\bibinfo  {journal} {Nature
  Communications}\ }\textbf {\bibinfo {volume} {2}},\ \bibinfo {pages} {396}
  (\bibinfo {year} {2011})}\BibitemShut {NoStop}%
\bibitem [{\citenamefont {Bangura}\ \emph {et~al.}(2013)\citenamefont
  {Bangura}, \citenamefont {Xu}, \citenamefont {Wakeham}, \citenamefont {Peng},
  \citenamefont {Horii},\ and\ \citenamefont {Hussey}}]{Bangura2013}%
  \BibitemOpen
  \bibfield  {author} {\bibinfo {author} {\bibfnamefont {A.~F.}\ \bibnamefont
  {Bangura}}, \bibinfo {author} {\bibfnamefont {X.}~\bibnamefont {Xu}},
  \bibinfo {author} {\bibfnamefont {N.}~\bibnamefont {Wakeham}}, \bibinfo
  {author} {\bibfnamefont {N.}~\bibnamefont {Peng}}, \bibinfo {author}
  {\bibfnamefont {S.}~\bibnamefont {Horii}}, \ and\ \bibinfo {author}
  {\bibfnamefont {N.~E.}\ \bibnamefont {Hussey}},\ }\href {\doibase
  10.1038/srep03261} {\bibfield  {journal} {\bibinfo  {journal} {Scientific
  Reports}\ }\textbf {\bibinfo {volume} {3}},\ \bibinfo {pages} {3261}
  (\bibinfo {year} {2013})}\BibitemShut {NoStop}%
\bibitem [{\citenamefont {Clarke}\ and\ \citenamefont
  {Strong}(1996)}]{ClarkeStrong1996}%
  \BibitemOpen
  \bibfield  {author} {\bibinfo {author} {\bibfnamefont {D.~G.}\ \bibnamefont
  {Clarke}}\ and\ \bibinfo {author} {\bibfnamefont {S.~P.}\ \bibnamefont
  {Strong}},\ }\href {\doibase 10.1080/00150199608216948} {\bibfield  {journal}
  {\bibinfo  {journal} {Ferroelectrics}\ }\textbf {\bibinfo {volume} {177}},\
  \bibinfo {pages} {1} (\bibinfo {year} {1996})}\BibitemShut {NoStop}%
\bibitem [{\citenamefont {Kane}\ and\ \citenamefont
  {Fisher}(1992{\natexlab{a}})}]{KaneFisher1992}%
  \BibitemOpen
  \bibfield  {author} {\bibinfo {author} {\bibfnamefont {C.~L.}\ \bibnamefont
  {Kane}}\ and\ \bibinfo {author} {\bibfnamefont {M.~P.~A.}\ \bibnamefont
  {Fisher}},\ }\href {\doibase 10.1103/PhysRevLett.68.1220} {\bibfield
  {journal} {\bibinfo  {journal} {Phys. Rev. Lett.}\ }\textbf {\bibinfo
  {volume} {68}},\ \bibinfo {pages} {1220} (\bibinfo {year}
  {1992}{\natexlab{a}})}\BibitemShut {NoStop}%
\bibitem [{\citenamefont {Kane}\ and\ \citenamefont
  {Fisher}(1992{\natexlab{b}})}]{KaneFisher1992b}%
  \BibitemOpen
  \bibfield  {author} {\bibinfo {author} {\bibfnamefont {C.~L.}\ \bibnamefont
  {Kane}}\ and\ \bibinfo {author} {\bibfnamefont {M.~P.~A.}\ \bibnamefont
  {Fisher}},\ }\href {\doibase 10.1103/PhysRevB.46.15233} {\bibfield  {journal}
  {\bibinfo  {journal} {Phys. Rev. B}\ }\textbf {\bibinfo {volume} {46}},\
  \bibinfo {pages} {15233} (\bibinfo {year} {1992}{\natexlab{b}})}\BibitemShut
  {NoStop}%
\bibitem [{\citenamefont {Kane}\ and\ \citenamefont
  {Fisher}(1996)}]{KaneFisher1996}%
  \BibitemOpen
  \bibfield  {author} {\bibinfo {author} {\bibfnamefont {C.~L.}\ \bibnamefont
  {Kane}}\ and\ \bibinfo {author} {\bibfnamefont {M.~P.~A.}\ \bibnamefont
  {Fisher}},\ }\href {\doibase 10.1103/PhysRevLett.76.3192} {\bibfield
  {journal} {\bibinfo  {journal} {Phys. Rev. Lett.}\ }\textbf {\bibinfo
  {volume} {76}},\ \bibinfo {pages} {3192} (\bibinfo {year}
  {1996})}\BibitemShut {NoStop}%
\bibitem [{\citenamefont {Li}\ and\ \citenamefont {Orignac}(2002)}]{Li2002}%
  \BibitemOpen
  \bibfield  {author} {\bibinfo {author} {\bibfnamefont {M.-R.}\ \bibnamefont
  {Li}}\ and\ \bibinfo {author} {\bibfnamefont {E.}~\bibnamefont {Orignac}},\
  }\href {http://stacks.iop.org/0295-5075/60/i=3/a=432} {\bibfield  {journal}
  {\bibinfo  {journal} {EPL (Europhysics Letters)}\ }\textbf {\bibinfo {volume}
  {60}},\ \bibinfo {pages} {432} (\bibinfo {year} {2002})}\BibitemShut
  {NoStop}%
\bibitem [{\citenamefont {Clarke}\ \emph {et~al.}(1995)\citenamefont {Clarke},
  \citenamefont {Strong},\ and\ \citenamefont {Anderson}}]{Anderson1995}%
  \BibitemOpen
  \bibfield  {author} {\bibinfo {author} {\bibfnamefont {D.~G.}\ \bibnamefont
  {Clarke}}, \bibinfo {author} {\bibfnamefont {S.~P.}\ \bibnamefont {Strong}},
  \ and\ \bibinfo {author} {\bibfnamefont {P.~W.}\ \bibnamefont {Anderson}},\
  }\href {\doibase 10.1103/PhysRevLett.74.4499} {\bibfield  {journal} {\bibinfo
   {journal} {Phys. Rev. Lett.}\ }\textbf {\bibinfo {volume} {74}},\ \bibinfo
  {pages} {4499} (\bibinfo {year} {1995})}\BibitemShut {NoStop}%
\bibitem [{\citenamefont {Georges}\ \emph {et~al.}(2000)\citenamefont
  {Georges}, \citenamefont {Giamarchi},\ and\ \citenamefont
  {Sandler}}]{Giamarchi2000_2013}%
  \BibitemOpen
  \bibfield  {author} {\bibinfo {author} {\bibfnamefont {A.}~\bibnamefont
  {Georges}}, \bibinfo {author} {\bibfnamefont {T.}~\bibnamefont {Giamarchi}},
  \ and\ \bibinfo {author} {\bibfnamefont {N.}~\bibnamefont {Sandler}},\ }\href
  {\doibase 10.1103/PhysRevB.61.16393} {\bibfield  {journal} {\bibinfo
  {journal} {Phys. Rev. B}\ }\textbf {\bibinfo {volume} {61}},\ \bibinfo
  {pages} {16393} (\bibinfo {year} {2000})}\BibitemShut {NoStop}%
\bibitem [{\citenamefont {Schattner}\ \emph {et~al.}(2016)\citenamefont
  {Schattner}, \citenamefont {Oganesyan},\ and\ \citenamefont
  {Orgad}}]{Schattner2016Arxiv}%
  \BibitemOpen
  \bibfield  {author} {\bibinfo {author} {\bibfnamefont {Y.}~\bibnamefont
  {Schattner}}, \bibinfo {author} {\bibfnamefont {V.}~\bibnamefont
  {Oganesyan}}, \ and\ \bibinfo {author} {\bibfnamefont {D.}~\bibnamefont
  {Orgad}},\ }\href {\doibase 10.1103/PhysRevB.94.235130} {\bibfield  {journal}
  {\bibinfo  {journal} {Phys. Rev. B}\ }\textbf {\bibinfo {volume} {94}},\
  \bibinfo {pages} {235130} (\bibinfo {year} {2016})}\BibitemShut {NoStop}%
\bibitem [{\citenamefont {Beni}(1974)}]{Beni1974}%
  \BibitemOpen
  \bibfield  {author} {\bibinfo {author} {\bibfnamefont {G.}~\bibnamefont
  {Beni}},\ }\href {\doibase 10.1103/PhysRevB.10.2186} {\bibfield  {journal}
  {\bibinfo  {journal} {Phys. Rev. B}\ }\textbf {\bibinfo {volume} {10}},\
  \bibinfo {pages} {2186} (\bibinfo {year} {1974})}\BibitemShut {NoStop}%
\bibitem [{\citenamefont {Mukerjee}(2005)}]{Mukerjee2005}%
  \BibitemOpen
  \bibfield  {author} {\bibinfo {author} {\bibfnamefont {S.}~\bibnamefont
  {Mukerjee}},\ }\href {\doibase 10.1103/PhysRevB.72.195109} {\bibfield
  {journal} {\bibinfo  {journal} {Phys. Rev. B}\ }\textbf {\bibinfo {volume}
  {72}},\ \bibinfo {pages} {195109} (\bibinfo {year} {2005})}\BibitemShut
  {NoStop}%
\bibitem [{Note1()}]{Note1}%
  \BibitemOpen
  \bibinfo {note} {Note that the term ``quasi-atomic limit'' has been used in
  the past to describe situations between full coherence and the atomic
  limit\cite {Siringo1991,Rojas2012}; we use it instead to indicate a system
  that is fully in the atomic limit in one direction and not at all in the
  other.}\BibitemShut {Stop}%
\bibitem [{\citenamefont {Clarke}\ \emph {et~al.}(1994)\citenamefont {Clarke},
  \citenamefont {Strong},\ and\ \citenamefont {Anderson}}]{Anderson1994}%
  \BibitemOpen
  \bibfield  {author} {\bibinfo {author} {\bibfnamefont {D.~G.}\ \bibnamefont
  {Clarke}}, \bibinfo {author} {\bibfnamefont {S.~P.}\ \bibnamefont {Strong}},
  \ and\ \bibinfo {author} {\bibfnamefont {P.~W.}\ \bibnamefont {Anderson}},\
  }\href {\doibase 10.1103/PhysRevLett.72.3218} {\bibfield  {journal} {\bibinfo
   {journal} {Phys. Rev. Lett.}\ }\textbf {\bibinfo {volume} {72}},\ \bibinfo
  {pages} {3218} (\bibinfo {year} {1994})}\BibitemShut {NoStop}%
\bibitem [{\citenamefont {Biermann}\ \emph {et~al.}(2002)\citenamefont
  {Biermann}, \citenamefont {Georges}, \citenamefont {Giamarchi},\ and\
  \citenamefont {Lichtenstein}}]{Biermann2002}%
  \BibitemOpen
  \bibfield  {author} {\bibinfo {author} {\bibfnamefont {S.}~\bibnamefont
  {Biermann}}, \bibinfo {author} {\bibfnamefont {A.}~\bibnamefont {Georges}},
  \bibinfo {author} {\bibfnamefont {T.}~\bibnamefont {Giamarchi}}, \ and\
  \bibinfo {author} {\bibfnamefont {A.}~\bibnamefont {Lichtenstein}},\
  }\enquote {\bibinfo {title} {Quasi one-dimensional organic conductors:
  Dimensional crossover and some puzzles},}\ in\ \href {\doibase
  10.1007/978-94-010-0530-2_5} {\emph {\bibinfo {booktitle} {Strongly
  Correlated Fermions and Bosons in Low-Dimensional Disordered Systems}}},\
  \bibinfo {editor} {edited by\ \bibinfo {editor} {\bibfnamefont {I.~V.}\
  \bibnamefont {Lerner}}, \bibinfo {editor} {\bibfnamefont {B.~L.}\
  \bibnamefont {Althsuler}}, \bibinfo {editor} {\bibfnamefont {V.~I.}\
  \bibnamefont {Fal'ko}}, \ and\ \bibinfo {editor} {\bibfnamefont
  {T.}~\bibnamefont {Giamarchi}}}\ (\bibinfo  {publisher} {Springer
  Netherlands},\ \bibinfo {address} {Dordrecht},\ \bibinfo {year} {2002})\ pp.\
  \bibinfo {pages} {81--102}\BibitemShut {NoStop}%
\bibitem [{\citenamefont {Noriega}\ \emph {et~al.}(2013)\citenamefont
  {Noriega}, \citenamefont {Rivnay}, \citenamefont {Vandewal}, \citenamefont
  {Koch}, \citenamefont {Stingelin}, \citenamefont {Smith}, \citenamefont
  {Toney},\ and\ \citenamefont {Salleo}}]{Stanford_experiment_paper}%
  \BibitemOpen
  \bibfield  {author} {\bibinfo {author} {\bibfnamefont {R.}~\bibnamefont
  {Noriega}}, \bibinfo {author} {\bibfnamefont {J.}~\bibnamefont {Rivnay}},
  \bibinfo {author} {\bibfnamefont {K.}~\bibnamefont {Vandewal}}, \bibinfo
  {author} {\bibfnamefont {F.~P.~V.}\ \bibnamefont {Koch}}, \bibinfo {author}
  {\bibfnamefont {N.}~\bibnamefont {Stingelin}}, \bibinfo {author}
  {\bibfnamefont {P.}~\bibnamefont {Smith}}, \bibinfo {author} {\bibfnamefont
  {M.~F.}\ \bibnamefont {Toney}}, \ and\ \bibinfo {author} {\bibfnamefont
  {A.}~\bibnamefont {Salleo}},\ }\href {\doibase 10.1038/nmat3722} {\bibfield
  {journal} {\bibinfo  {journal} {Nat Mater}\ }\textbf {\bibinfo {volume}
  {12}},\ \bibinfo {pages} {1038} (\bibinfo {year} {2013})}\BibitemShut
  {NoStop}%
\bibitem [{\citenamefont {Steyrleuthner}\ \emph {et~al.}(2014)\citenamefont
  {Steyrleuthner}, \citenamefont {Di~Pietro}, \citenamefont {Collins},
  \citenamefont {Polzer}, \citenamefont {Himmelberger}, \citenamefont
  {Schubert}, \citenamefont {Chen}, \citenamefont {Zhang}, \citenamefont
  {Salleo}, \citenamefont {Ade}, \citenamefont {Facchetti},\ and\ \citenamefont
  {Neher}}]{Steyrleuthner2014}%
  \BibitemOpen
  \bibfield  {author} {\bibinfo {author} {\bibfnamefont {R.}~\bibnamefont
  {Steyrleuthner}}, \bibinfo {author} {\bibfnamefont {R.}~\bibnamefont
  {Di~Pietro}}, \bibinfo {author} {\bibfnamefont {B.~A.}\ \bibnamefont
  {Collins}}, \bibinfo {author} {\bibfnamefont {F.}~\bibnamefont {Polzer}},
  \bibinfo {author} {\bibfnamefont {S.}~\bibnamefont {Himmelberger}}, \bibinfo
  {author} {\bibfnamefont {M.}~\bibnamefont {Schubert}}, \bibinfo {author}
  {\bibfnamefont {Z.}~\bibnamefont {Chen}}, \bibinfo {author} {\bibfnamefont
  {S.}~\bibnamefont {Zhang}}, \bibinfo {author} {\bibfnamefont
  {A.}~\bibnamefont {Salleo}}, \bibinfo {author} {\bibfnamefont
  {H.}~\bibnamefont {Ade}}, \bibinfo {author} {\bibfnamefont {A.}~\bibnamefont
  {Facchetti}}, \ and\ \bibinfo {author} {\bibfnamefont {D.}~\bibnamefont
  {Neher}},\ }\href {\doibase 10.1021/ja4118736} {\bibfield  {journal}
  {\bibinfo  {journal} {Journal of the American Chemical Society}\ }\textbf
  {\bibinfo {volume} {136}},\ \bibinfo {pages} {4245} (\bibinfo {year}
  {2014})}\BibitemShut {NoStop}%
\bibitem [{\citenamefont {Mahan}(2000)}]{Mahan3}%
  \BibitemOpen
  \bibfield  {author} {\bibinfo {author} {\bibfnamefont {G.~D.}\ \bibnamefont
  {Mahan}},\ }\href@noop {} {\emph {\bibinfo {title} {Many-Particle Physics --
  3rd ed.}}}\ (\bibinfo  {publisher} {Springer Science + Business Media},\
  \bibinfo {address} {New York},\ \bibinfo {year} {2000})\BibitemShut {NoStop}%
\bibitem [{Note2()}]{Note2}%
  \BibitemOpen
  \bibinfo {note} {Equation \protect \textup {\hbox {\mathsurround \z@ \protect
  \normalfont (\ignorespaces \ref {eq:L_Kubo_formula}\unskip \@@italiccorr )}}
  is a corrected version of (3.518) from reference \protect \rev@citealpnum
  {Mahan3}; see the Supplemental Material\cite {SuppMat} for
  details.}\BibitemShut {Stop}%
\bibitem [{\citenamefont {Mattsson}\ \emph {et~al.}(1997)\citenamefont
  {Mattsson}, \citenamefont {Eggert},\ and\ \citenamefont
  {Johannesson}}]{finiteLLpaper}%
  \BibitemOpen
  \bibfield  {author} {\bibinfo {author} {\bibfnamefont {A.~E.}\ \bibnamefont
  {Mattsson}}, \bibinfo {author} {\bibfnamefont {S.}~\bibnamefont {Eggert}}, \
  and\ \bibinfo {author} {\bibfnamefont {H.}~\bibnamefont {Johannesson}},\
  }\href {\doibase 10.1103/PhysRevB.56.15615} {\bibfield  {journal} {\bibinfo
  {journal} {Phys. Rev. B}\ }\textbf {\bibinfo {volume} {56}},\ \bibinfo
  {pages} {15615} (\bibinfo {year} {1997})}\BibitemShut {NoStop}%
\bibitem [{\citenamefont {Blagouchine}(2015)}]{Blagouchine2014}%
  \BibitemOpen
  \bibfield  {author} {\bibinfo {author} {\bibfnamefont {I.~V.}\ \bibnamefont
  {Blagouchine}},\ }\href {\doibase 10.1016/j.jnt.2014.08.009} {\bibfield
  {journal} {\bibinfo  {journal} {Journal of Number Theory}\ }\textbf {\bibinfo
  {volume} {148}},\ \bibinfo {pages} {537} (\bibinfo {year}
  {2015})}\BibitemShut {NoStop}%
\bibitem [{Note3()}]{Note3}%
  \BibitemOpen
  \bibinfo {note} {For purposes of calculation, the generalized Stieltjes
  constant is implemented in the commercial software Wolfram Mathematica as
  $\gamma _n(\nu )=\protect \text {StieltjesGamma}[n,\nu ]$\cite
  {WolframSGDef}}\BibitemShut {NoStop}%
\bibitem [{{\relax DLMF}()}]{NIST:DLMF}%
  \BibitemOpen
  {\relax DLMF},\ \href {http://dlmf.nist.gov/} {\enquote {\bibinfo {title}
  {{\it NIST Digital Library of Mathematical Functions}},}\ }\bibinfo
  {howpublished} {http://dlmf.nist.gov/, Release 1.0.13 of 2016-09-16},\
  \bibinfo {note} {F.~W.~J. Olver, A.~B. {Olde Daalhuis}, D.~W. Lozier, B.~I.
  Schneider, R.~F. Boisvert, C.~W. Clark, B.~R. Miller and B.~V. Saunders,
  eds.}\BibitemShut {Stop}%
\bibitem [{Note4()}]{Note4}%
  \BibitemOpen
  \bibinfo {note} {For purposes of calculation, the function $F_1$ is
  implemented in the commercial software Wolfram Mathematica as
  $F_1(a;b_1,b_2;c;x,y)=\protect \text {AppellF1}[a,b1,b2,c,x,y]$\cite
  {WolframAppellDef}}\BibitemShut {NoStop}%
\bibitem [{Sup()}]{SuppMat}%
  \BibitemOpen
  \href@noop {} {}\bibinfo {note} {See \href{http://cmt.berkeley.edu/suppl/szasz-arxiv16-suppl.pdf}{Supplemental Material} for details of the calculations.}\BibitemShut {Stop}%
\bibitem [{\citenamefont {Vescoli}\ \emph {et~al.}(1998)\citenamefont
  {Vescoli}, \citenamefont {Degiorgi}, \citenamefont {Henderson}, \citenamefont
  {Gr{\"u}ner}, \citenamefont {Starkey},\ and\ \citenamefont
  {Montgomery}}]{Vescoli1998}%
  \BibitemOpen
  \bibfield  {author} {\bibinfo {author} {\bibfnamefont {V.}~\bibnamefont
  {Vescoli}}, \bibinfo {author} {\bibfnamefont {L.}~\bibnamefont {Degiorgi}},
  \bibinfo {author} {\bibfnamefont {W.}~\bibnamefont {Henderson}}, \bibinfo
  {author} {\bibfnamefont {G.}~\bibnamefont {Gr{\"u}ner}}, \bibinfo {author}
  {\bibfnamefont {K.~P.}\ \bibnamefont {Starkey}}, \ and\ \bibinfo {author}
  {\bibfnamefont {L.~K.}\ \bibnamefont {Montgomery}},\ }\href {\doibase
  10.1126/science.281.5380.1181} {\bibfield  {journal} {\bibinfo  {journal}
  {Science}\ }\textbf {\bibinfo {volume} {281}},\ \bibinfo {pages} {1181}
  (\bibinfo {year} {1998})}\BibitemShut {NoStop}%
\bibitem [{\citenamefont {Schulz}(1990)}]{Schulz1990}%
  \BibitemOpen
  \bibfield  {author} {\bibinfo {author} {\bibfnamefont {H.~J.}\ \bibnamefont
  {Schulz}},\ }\href {\doibase 10.1103/PhysRevLett.64.2831} {\bibfield
  {journal} {\bibinfo  {journal} {Phys. Rev. Lett.}\ }\textbf {\bibinfo
  {volume} {64}},\ \bibinfo {pages} {2831} (\bibinfo {year}
  {1990})}\BibitemShut {NoStop}%
\bibitem [{\citenamefont {Ejima}\ \emph {et~al.}(2005)\citenamefont {Ejima},
  \citenamefont {Gebhard},\ and\ \citenamefont {Nishimoto}}]{Ejima2005}%
  \BibitemOpen
  \bibfield  {author} {\bibinfo {author} {\bibfnamefont {S.}~\bibnamefont
  {Ejima}}, \bibinfo {author} {\bibfnamefont {F.}~\bibnamefont {Gebhard}}, \
  and\ \bibinfo {author} {\bibfnamefont {S.}~\bibnamefont {Nishimoto}},\ }\href
  {http://stacks.iop.org/0295-5075/70/i=4/a=492} {\bibfield  {journal}
  {\bibinfo  {journal} {EPL (Europhysics Letters)}\ }\textbf {\bibinfo {volume}
  {70}},\ \bibinfo {pages} {492} (\bibinfo {year} {2005})}\BibitemShut
  {NoStop}%
\bibitem [{\citenamefont {Ejima}\ \emph {et~al.}(2006)\citenamefont {Ejima},
  \citenamefont {Gebhard},\ and\ \citenamefont {Nishimoto}}]{Ejima2006}%
  \BibitemOpen
  \bibfield  {author} {\bibinfo {author} {\bibfnamefont {S.}~\bibnamefont
  {Ejima}}, \bibinfo {author} {\bibfnamefont {F.}~\bibnamefont {Gebhard}}, \
  and\ \bibinfo {author} {\bibfnamefont {S.}~\bibnamefont {Nishimoto}},\ }\href
  {\doibase 10.1103/PhysRevB.74.245110} {\bibfield  {journal} {\bibinfo
  {journal} {Phys. Rev. B}\ }\textbf {\bibinfo {volume} {74}},\ \bibinfo
  {pages} {245110} (\bibinfo {year} {2006})}\BibitemShut {NoStop}%
\bibitem [{\citenamefont {Sirker}(2012)}]{Sirker2012}%
  \BibitemOpen
  \bibfield  {author} {\bibinfo {author} {\bibfnamefont {J.}~\bibnamefont
  {Sirker}},\ }\href {\doibase 10.1142/S0217979212440092} {\bibfield  {journal}
  {\bibinfo  {journal} {International Journal of Modern Physics B}\ }\textbf
  {\bibinfo {volume} {26}},\ \bibinfo {pages} {1244009} (\bibinfo {year}
  {2012})}\BibitemShut {NoStop}%
\bibitem [{\citenamefont {Affleck}(2005)}]{Affleck2013}%
  \BibitemOpen
  \bibfield  {author} {\bibinfo {author} {\bibfnamefont {I.}~\bibnamefont
  {Affleck}},\ }\href {\doibase 10.1103/PhysRevB.72.132414} {\bibfield
  {journal} {\bibinfo  {journal} {Phys. Rev. B}\ }\textbf {\bibinfo {volume}
  {72}},\ \bibinfo {pages} {132414} (\bibinfo {year} {2005})}\BibitemShut
  {NoStop}%
\bibitem [{\citenamefont {Isichenko}(1992)}]{Isichenko1992}%
  \BibitemOpen
  \bibfield  {author} {\bibinfo {author} {\bibfnamefont {M.~B.}\ \bibnamefont
  {Isichenko}},\ }\href {\doibase 10.1103/RevModPhys.64.961} {\bibfield
  {journal} {\bibinfo  {journal} {Rev. Mod. Phys.}\ }\textbf {\bibinfo {volume}
  {64}},\ \bibinfo {pages} {961} (\bibinfo {year} {1992})}\BibitemShut
  {NoStop}%
\bibitem [{\citenamefont {Vidal}(2003)}]{vidalrg}%
  \BibitemOpen
  \bibfield  {author} {\bibinfo {author} {\bibfnamefont {G.}~\bibnamefont
  {Vidal}},\ }\href {\doibase 10.1103/PhysRevLett.91.147902} {\bibfield
  {journal} {\bibinfo  {journal} {Phys. Rev. Lett.}\ }\textbf {\bibinfo
  {volume} {91}},\ \bibinfo {pages} {147902} (\bibinfo {year}
  {2003})}\BibitemShut {NoStop}%
\bibitem [{\citenamefont {White}\ and\ \citenamefont
  {Feiguin}(2004)}]{whitetdmrg}%
  \BibitemOpen
  \bibfield  {author} {\bibinfo {author} {\bibfnamefont {S.~R.}\ \bibnamefont
  {White}}\ and\ \bibinfo {author} {\bibfnamefont {A.~E.}\ \bibnamefont
  {Feiguin}},\ }\href {\doibase 10.1103/PhysRevLett.93.076401} {\bibfield
  {journal} {\bibinfo  {journal} {Phys. Rev. Lett.}\ }\textbf {\bibinfo
  {volume} {93}},\ \bibinfo {pages} {076401} (\bibinfo {year}
  {2004})}\BibitemShut {NoStop}%
\bibitem [{\citenamefont {Schollwoeck}(2011)}]{schollwoeck}%
  \BibitemOpen
  \bibfield  {author} {\bibinfo {author} {\bibfnamefont {U.}~\bibnamefont
  {Schollwoeck}},\ }\href {\doibase
  http://dx.doi.org/10.1016/j.aop.2010.09.012} {\bibfield  {journal} {\bibinfo
  {journal} {Annals of Physics}\ }\textbf {\bibinfo {volume} {326}},\ \bibinfo
  {pages} {96 } (\bibinfo {year} {2011})},\ \bibinfo {note} {january 2011
  Special Issue}\BibitemShut {NoStop}%
\bibitem [{\citenamefont {Huang}\ \emph {et~al.}(2013)\citenamefont {Huang},
  \citenamefont {Karrasch},\ and\ \citenamefont {Moore}}]{huangkarrasch}%
  \BibitemOpen
  \bibfield  {author} {\bibinfo {author} {\bibfnamefont {Y.}~\bibnamefont
  {Huang}}, \bibinfo {author} {\bibfnamefont {C.}~\bibnamefont {Karrasch}}, \
  and\ \bibinfo {author} {\bibfnamefont {J.~E.}\ \bibnamefont {Moore}},\ }\href
  {\doibase 10.1103/PhysRevB.88.115126} {\bibfield  {journal} {\bibinfo
  {journal} {Phys. Rev. B}\ }\textbf {\bibinfo {volume} {88}},\ \bibinfo
  {pages} {115126} (\bibinfo {year} {2013})}\BibitemShut {NoStop}%
\bibitem [{\citenamefont {James}\ and\ \citenamefont
  {Konik}(2013)}]{jameskonik}%
  \BibitemOpen
  \bibfield  {author} {\bibinfo {author} {\bibfnamefont {A.~J.~A.}\
  \bibnamefont {James}}\ and\ \bibinfo {author} {\bibfnamefont {R.~M.}\
  \bibnamefont {Konik}},\ }\href {\doibase 10.1103/PhysRevB.87.241103}
  {\bibfield  {journal} {\bibinfo  {journal} {Phys. Rev. B}\ }\textbf {\bibinfo
  {volume} {87}},\ \bibinfo {pages} {241103} (\bibinfo {year}
  {2013})}\BibitemShut {NoStop}%
\bibitem [{\citenamefont {Jarrell}\ and\ \citenamefont
  {Gubernatis}(1996)}]{Jarrell1996}%
  \BibitemOpen
  \bibfield  {author} {\bibinfo {author} {\bibfnamefont {M.}~\bibnamefont
  {Jarrell}}\ and\ \bibinfo {author} {\bibfnamefont {J.}~\bibnamefont
  {Gubernatis}},\ }\href {\doibase
  http://dx.doi.org/10.1016/0370-1573(95)00074-7} {\bibfield  {journal}
  {\bibinfo  {journal} {Physics Reports}\ }\textbf {\bibinfo {volume} {269}},\
  \bibinfo {pages} {133 } (\bibinfo {year} {1996})}\BibitemShut {NoStop}%
\bibitem [{\citenamefont {Gunnarsson}\ \emph {et~al.}(2010)\citenamefont
  {Gunnarsson}, \citenamefont {Haverkort},\ and\ \citenamefont
  {Sangiovanni}}]{Gunnarsson2010}%
  \BibitemOpen
  \bibfield  {author} {\bibinfo {author} {\bibfnamefont {O.}~\bibnamefont
  {Gunnarsson}}, \bibinfo {author} {\bibfnamefont {M.~W.}\ \bibnamefont
  {Haverkort}}, \ and\ \bibinfo {author} {\bibfnamefont {G.}~\bibnamefont
  {Sangiovanni}},\ }\href {\doibase 10.1103/PhysRevB.81.155107} {\bibfield
  {journal} {\bibinfo  {journal} {Phys. Rev. B}\ }\textbf {\bibinfo {volume}
  {81}},\ \bibinfo {pages} {155107} (\bibinfo {year} {2010})}\BibitemShut
  {NoStop}%
\bibitem [{\citenamefont {Vidberg}\ and\ \citenamefont
  {Serene}(1977)}]{Vidberg1977}%
  \BibitemOpen
  \bibfield  {author} {\bibinfo {author} {\bibfnamefont {H.~J.}\ \bibnamefont
  {Vidberg}}\ and\ \bibinfo {author} {\bibfnamefont {J.~W.}\ \bibnamefont
  {Serene}},\ }\href {\doibase 10.1007/BF00655090} {\bibfield  {journal}
  {\bibinfo  {journal} {Journal of Low Temperature Physics}\ }\textbf {\bibinfo
  {volume} {29}},\ \bibinfo {pages} {179} (\bibinfo {year} {1977})}\BibitemShut
  {NoStop}%
\bibitem [{\citenamefont {Osolin}\ and\ \citenamefont
  {\v{Z}itko}(2013)}]{Pade_ref}%
  \BibitemOpen
  \bibfield  {author} {\bibinfo {author} {\bibfnamefont {\v{Z}.}~\bibnamefont
  {Osolin}}\ and\ \bibinfo {author} {\bibfnamefont {R.}~\bibnamefont
  {\v{Z}itko}},\ }\href {\doibase 10.1103/PhysRevB.87.245135} {\bibfield
  {journal} {\bibinfo  {journal} {Phys. Rev. B}\ }\textbf {\bibinfo {volume}
  {87}},\ \bibinfo {pages} {245135} (\bibinfo {year} {2013})}\BibitemShut
  {NoStop}%
\bibitem [{\citenamefont {Siringo}\ and\ \citenamefont
  {Logan}(1991)}]{Siringo1991}%
  \BibitemOpen
  \bibfield  {author} {\bibinfo {author} {\bibfnamefont {F.}~\bibnamefont
  {Siringo}}\ and\ \bibinfo {author} {\bibfnamefont {D.~E.}\ \bibnamefont
  {Logan}},\ }\href {http://stacks.iop.org/0953-8984/3/i=25/a=020} {\bibfield
  {journal} {\bibinfo  {journal} {Journal of Physics: Condensed Matter}\
  }\textbf {\bibinfo {volume} {3}},\ \bibinfo {pages} {4747} (\bibinfo {year}
  {1991})}\BibitemShut {NoStop}%
\bibitem [{\citenamefont {Rojas}\ \emph {et~al.}(2012)\citenamefont {Rojas},
  \citenamefont {de~Souza},\ and\ \citenamefont {Ananikian}}]{Rojas2012}%
  \BibitemOpen
  \bibfield  {author} {\bibinfo {author} {\bibfnamefont {O.}~\bibnamefont
  {Rojas}}, \bibinfo {author} {\bibfnamefont {S.~M.}\ \bibnamefont {de~Souza}},
  \ and\ \bibinfo {author} {\bibfnamefont {N.~S.}\ \bibnamefont {Ananikian}},\
  }\href {\doibase 10.1103/PhysRevE.85.061123} {\bibfield  {journal} {\bibinfo
  {journal} {Phys. Rev. E}\ }\textbf {\bibinfo {volume} {85}},\ \bibinfo
  {pages} {061123} (\bibinfo {year} {2012})}\BibitemShut {NoStop}%
\bibitem [{\citenamefont {Weisstein}({\natexlab{a}})}]{WolframSGDef}%
  \BibitemOpen
  \bibfield  {author} {\bibinfo {author} {\bibfnamefont {E.~W.}\ \bibnamefont
  {Weisstein}},\ }\href@noop {} {\enquote {\bibinfo {title} {Stieltjes
  constants},}\ }\bibinfo {howpublished} {from \textit{MathWorld}--A Wolfram
  Web Resource. \href{http://mathworld.wolfram.com/StieltjesConstants.html}{http://mathworld.wolfram.com/StieltjesConstants.html}}  
  \BibitemShut {NoStop}%
\bibitem [{\citenamefont {Weisstein}({\natexlab{b}})}]{WolframAppellDef}%
  \BibitemOpen
  \bibfield  {author} {\bibinfo {author} {\bibfnamefont {E.~W.}\ \bibnamefont
  {Weisstein}},\ }\href@noop {} {\enquote {\bibinfo {title} {Appell
  hypergeometric function},}\ }\bibinfo {howpublished} {from
  \textit{MathWorld}--A Wolfram Web Resource.
  \href{http://mathworld.wolfram.com/AppellHypergeometricFunction.html}{http://mathworld.wolfram.com/AppellHypergeometric}\\ \href{http://mathworld.wolfram.com/AppellHypergeometricFunction.html}{Function.html}}
  \BibitemShut {NoStop}%
\end{thebibliography}
\end{document}